\documentclass[aps,pra,reprint,superscriptaddress]{revtex4-1}
\usepackage{graphicx}
\usepackage{fullpage}
\usepackage{upgreek}
\usepackage{verbatim}

\begin{document}

\title{Spectroscopy of Ba and Ba$^+$ deposits in solid xenon for barium tagging in nEXO} 

\author{B.~Mong}
\affiliation{Physics Department, Colorado State University, Fort Collins CO, USA}
\affiliation{Department of Physics, Laurentian University, Sudbury ON, Canada}
\author{S.~Cook}\thanks{Now at NIST, Boulder CO, USA}
\affiliation{Physics Department, Colorado State University, Fort Collins CO, USA}
\author{T.~Walton}
\affiliation{Physics Department, Colorado State University, Fort Collins CO, USA}
\author{C.~Chambers}
\affiliation{Physics Department, Colorado State University, Fort Collins CO, USA}
\author{A.~Craycraft}
\affiliation{Physics Department, Colorado State University, Fort Collins CO, USA}
\author{C.~Benitez-Medina}\thanks{Now at Intel, Hillsboro OR, USA}
\affiliation{Physics Department, Colorado State University, Fort Collins CO, USA}
\author{K.~Hall}\thanks{Now at Raytheon, Denver CO, USA}
\affiliation{Physics Department, Colorado State University, Fort Collins CO, USA}
\author{W.~Fairbank Jr.}\thanks{Corresponding author}
\affiliation{Physics Department, Colorado State University, Fort Collins CO, USA}
\author{J.B.~Albert}
\affiliation{Physics~Department~and~CEEM,~Indiana~University,~Bloomington~IN,~USA}
\author{D.J.~Auty}
\affiliation{Department of Physics and Astronomy, University of Alabama, Tuscaloosa AL, USA}
\author{P.S.~Barbeau}
\affiliation{Department of Physics, Duke University, and Triangle Universities Nuclear Laboratory (TUNL), Durham North Carolina, USA}
\author{ V. ~Basque}
\affiliation{Physics Department, Carleton University, Ottawa ON, Canada}
\author{D.~Beck}
\affiliation{Physics Department, University of Illinois, Urbana-Champaign IL, USA}
\author{M.~Breidenbach}
\affiliation{SLAC National Accelerator Laboratory, Stanford CA, USA}
\author{T.~Brunner}
\affiliation{Physics Department, Stanford University, Stanford CA, USA}
\author{G.F.~Cao}
\affiliation{Institute of High Energy Physics, Beijing, China}
\author{B.~Cleveland}\thanks{Also SNOLAB, Sudbury ON, Canada}
\affiliation{Department of Physics, Laurentian University, Sudbury ON, Canada}
\author{M.~Coon}
\affiliation{Physics Department, University of Illinois, Urbana-Champaign IL, USA}
\author{T.~Daniels}
\affiliation{SLAC National Accelerator Laboratory, Stanford CA, USA}
\author{S.J.~Daugherty}
\affiliation{Physics~Department~and~CEEM,~Indiana~University,~Bloomington~IN,~USA}
\author{R.~DeVoe}
\affiliation{Physics Department, Stanford University, Stanford CA, USA}
\author{T.~Didberidze}
\affiliation{Department of Physics and Astronomy, University of Alabama, Tuscaloosa AL, USA}
\author{J.~Dilling}
\affiliation{TRIUMF, Vancouver BC, Canada}
\author{M.J.~Dolinski}
\affiliation{Department of Physics, Drexel University, Philadelphia PA, USA}
\author{M.~Dunford}
\affiliation{Physics Department, Carleton University, Ottawa ON, Canada}
\author{L.~Fabris}
\affiliation{Oak Ridge National Laboratory, Oak Ridge TN, USA}
\author{J.~Farine}
\affiliation{Department of Physics, Laurentian University, Sudbury ON, Canada}
\author{W.~Feldmeier}
\affiliation{Technische Universitat Munchen, Physikdepartment and Excellence Cluster Universe, Garching, Germany}
\author{P.~Fierlinger}
\affiliation{Technische Universitat Munchen, Physikdepartment and Excellence Cluster Universe, Garching, Germany}
\author{D.~Fudenberg}
\affiliation{Physics Department, Stanford University, Stanford CA, USA}
\author{G.~Giroux}\thanks{Now at Queen's University, Kingston ON, Canada}
\affiliation{LHEP, Albert Einstein Center, University of Bern, Bern, Switzerland}
\author{R.~Gornea}
\affiliation{LHEP, Albert Einstein Center, University of Bern, Bern, Switzerland}
\author{K.~Graham}
\affiliation{Physics Department, Carleton University, Ottawa ON, Canada}
\author{G.~Gratta}
\affiliation{Physics Department, Stanford University, Stanford CA, USA}
\author{M.~Heffner}
\affiliation{Lawrence Livermore National Laboratory, Livermore CA, USA}
\author{M.~Hughes}
\affiliation{Department of Physics and Astronomy, University of Alabama, Tuscaloosa AL, USA}
\author{X.S.~Jiang}
\affiliation{Institute of High Energy Physics, Beijing, China}
\author{T.N.~Johnson}
\affiliation{Physics~Department~and~CEEM,~Indiana~University,~Bloomington~IN,~USA}
\author{S.~Johnston}
\affiliation{Physics Department, University of Massachusetts, Amherst MA, USA}
\author{A.~Karelin}
\affiliation{Institute for Theoretical and Experimental Physics, Moscow, Russia}
\author{L.J.~Kaufman}
\affiliation{Physics~Department~and~CEEM,~Indiana~University,~Bloomington~IN,~USA}
\author{R.~Killick}
\affiliation{Physics Department, Carleton University, Ottawa ON, Canada}
\author{T.~Koffas}
\affiliation{Physics Department, Carleton University, Ottawa ON, Canada}
\author{S.~Kravitz}
\affiliation{Physics Department, Stanford University, Stanford CA, USA}
\author{R.~Kr\"ucken}
\affiliation{TRIUMF, Vancouver BC, Canada}
\author{A.~Kuchenkov}
\affiliation{Institute for Theoretical and Experimental Physics, Moscow, Russia}
\author{K.S.~Kumar}
\affiliation{Department of Physics and Astronomy, Stony Brook University, SUNY, Stony Brook NY,USA}
\author{D.S.~Leonard}
\affiliation{Department of Physics, University of Seoul, Seoul, Korea}
\author{C.~Licciardi}
\affiliation{Physics Department, Carleton University, Ottawa ON, Canada}
\author{Y.H.~Lin}
\affiliation{Department of Physics, Drexel University, Philadelphia PA, USA}
\author{J.~Ling}
\affiliation{Physics Department, University of Illinois, Urbana-Champaign IL, USA}
\author{R.~MacLellan}
\affiliation{Department of Physics, University of South Dakota, Vermillion SD, USA}
\author{M.G.~Marino}
\affiliation{Technische Universitat Munchen, Physikdepartment and Excellence Cluster Universe, Garching, Germany}
\author{D.~Moore}
\affiliation{Physics Department, Stanford University, Stanford CA, USA}
\author{A.~Odian}
\affiliation{SLAC National Accelerator Laboratory, Stanford CA, USA}
\author{I.~Ostrovskiy}
\affiliation{Physics Department, Stanford University, Stanford CA, USA}
\author{A.~Piepke}
\affiliation{Department of Physics and Astronomy, University of Alabama, Tuscaloosa AL, USA}
\author{A.~Pocar}
\affiliation{Physics Department, University of Massachusetts, Amherst MA, USA}
\author{F.~Retiere}
\affiliation{TRIUMF, Vancouver BC, Canada}
\author{P.C.~Rowson}
\affiliation{SLAC National Accelerator Laboratory, Stanford CA, USA}
\author{M.P.~Rozo}
\affiliation{Physics Department, Carleton University, Ottawa ON, Canada}
\author{A.~Schubert}
\affiliation{Physics Department, Stanford University, Stanford CA, USA}
\author{D.~Sinclair}
\affiliation{TRIUMF, Vancouver BC, Canada}
\affiliation{Physics Department, Carleton University, Ottawa ON, Canada}
\author{E.~Smith}
\affiliation{Department of Physics, Drexel University, Philadelphia PA, USA}
\author{V.~Stekhanov}
\affiliation{Institute for Theoretical and Experimental Physics, Moscow, Russia}
\author{M.~Tarka}
\affiliation{Physics Department, University of Illinois, Urbana-Champaign IL, USA}
\author{T.~Tolba}
\affiliation{LHEP, Albert Einstein Center, University of Bern, Bern, Switzerland}
\author{K.~Twelker}
\affiliation{Physics Department, Stanford University, Stanford CA, USA}
\author{J.-L.~Vuilleumier}
\affiliation{LHEP, Albert Einstein Center, University of Bern, Bern, Switzerland}
\author{J.~Walton}
\affiliation{Physics Department, University of Illinois, Urbana-Champaign IL, USA}
\author{M.~Weber}
\affiliation{Physics Department, Stanford University, Stanford CA, USA}
\author{L.J.~Wen}
\affiliation{Institute of High Energy Physics, Beijing, China}
\author{U.~Wichoski}
\affiliation{Department of Physics, Laurentian University, Sudbury ON, Canada}
\author{L.~Yang}
\affiliation{Physics Department, University of Illinois, Urbana-Champaign IL, USA}
\author{Y.-R.~Yen}
\affiliation{Department of Physics, Drexel University, Philadelphia PA, USA}
\author{Y.B.~Zhao}
\affiliation{Institute of High Energy Physics, Beijing, China}

\date{\today}

\begin{abstract}
Progress on a method of barium tagging for the nEXO double beta decay experiment is reported.  
Absorption and emission spectra for deposits of barium atoms and ions in solid xenon matrices are presented.
Excitation spectra for prominent emission lines, temperature dependence and bleaching of the fluorescence reveal the existence of different matrix sites.  
A regular series of sharp lines observed in Ba$^+$ deposits is identified with some type of barium hydride molecule.  
Lower limits for the fluorescence quantum efficiency of the principal Ba emission transition are reported.
Under current conditions, an image of $\le10^4$ Ba atoms can be obtained.
Prospects for imaging single Ba atoms in solid xenon are discussed.

\end{abstract}


\maketitle 

\section{Introduction}

The spectroscopy of atoms and molecules isolated in solid matrices of inert gases dates back sixty years \cite{Whittle1954}.
Matrix isolation spectroscopy, as this method is known, has established that atomic states in noble gas matrices retain many of the fundamental properties of their vacuum counterparts, such as the quantum numbers, even in the most polarizable noble gas matrix, solid xenon (SXe) \cite{Crepin1999}.
Peaks in the condensed phase spectra, though broadened to nanometers in width and somewhat red- or blue-shifted, can be assigned to known transitions of free atoms in simple cases such as alkali, alkaline earth and transition metal atoms.
As the field has matured, the ability to capture, hold and study small numbers of atoms for a long period of time has attracted interest for difficult and sensitive applications.

A novel application of matrix isolation spectroscopy is being explored by the nEXO collaboration for a future ton-scale $ ^{136} $Xe neutrinoless double beta decay ($ 0\nu\beta\beta $) experiment.
Successful observation of $ 0\nu\beta\beta $ decay would determine the fundamental character of neutrinos to be Majorana rather than Dirac, and could provide the additional information needed to infer the absolute masses of the neutrinos \cite{Avignone2008}.
The most sensitive $ 0\nu\beta\beta $ decay experiments to date, with tens to hundreds of kilograms of the isotope of interest, have reached $ 0\nu\beta\beta $ decay lifetime limits of greater than $ 10^{25} $ years \cite{Auger2012,Gando2013,Agostini2013,Albert2014}.
In next generation experiments at the ton scale, complete elimination of background would be a great advantage, as the sensitivity to lifetime grows linearly with the mass of the isotope in the detector, M, in the zero background case, whereas the sensitivity grows as M$^{1/2} $ with backgrounds that increase proportionately with mass.  

Among all the double beta decay isotopes, $ ^{136} $Xe is unique because the decay medium can be a transparent liquid or gas.
The $ 0\nu\beta\beta $ decay of $ ^{136}$Xe produces a daughter ion $ ^{136} $Ba$^{++}$ ($^{136}$Xe $\rightarrow^{136}$Ba$^{++} + 2e^{-}$). 
In liquid xenon (LXe), it has been proposed that after single charge transfer in the liquid, the daughter $ ^{136}$Ba$^+$ ion might be identified $in~situ$ by laser spectroscopy at the site of the decay.
With this additional identification, or ``tag'', all backgrounds of a $^{136}$Xe  $0\nu\beta\beta $ decay experiment in the energy range of interest near the Q value could be vetoed.
The one exception, background from the two-neutrino double beta decays, is estimated to be negligible still for multi-ton detectors and observed energy resolutions.
The original barium tagging proposal  \cite{Moe1991} called for exciting and detecting the $^{136}$Ba ion by lasers directed through the liquid xenon to the decay site.
However, our efforts to demonstrate direct barium tagging in liquid xenon with lasers have thus far been inconclusive \cite{Hall2012}.

\begin{figure}[h!tb]
	\begin{center}
	\includegraphics[width=0.45\textwidth]{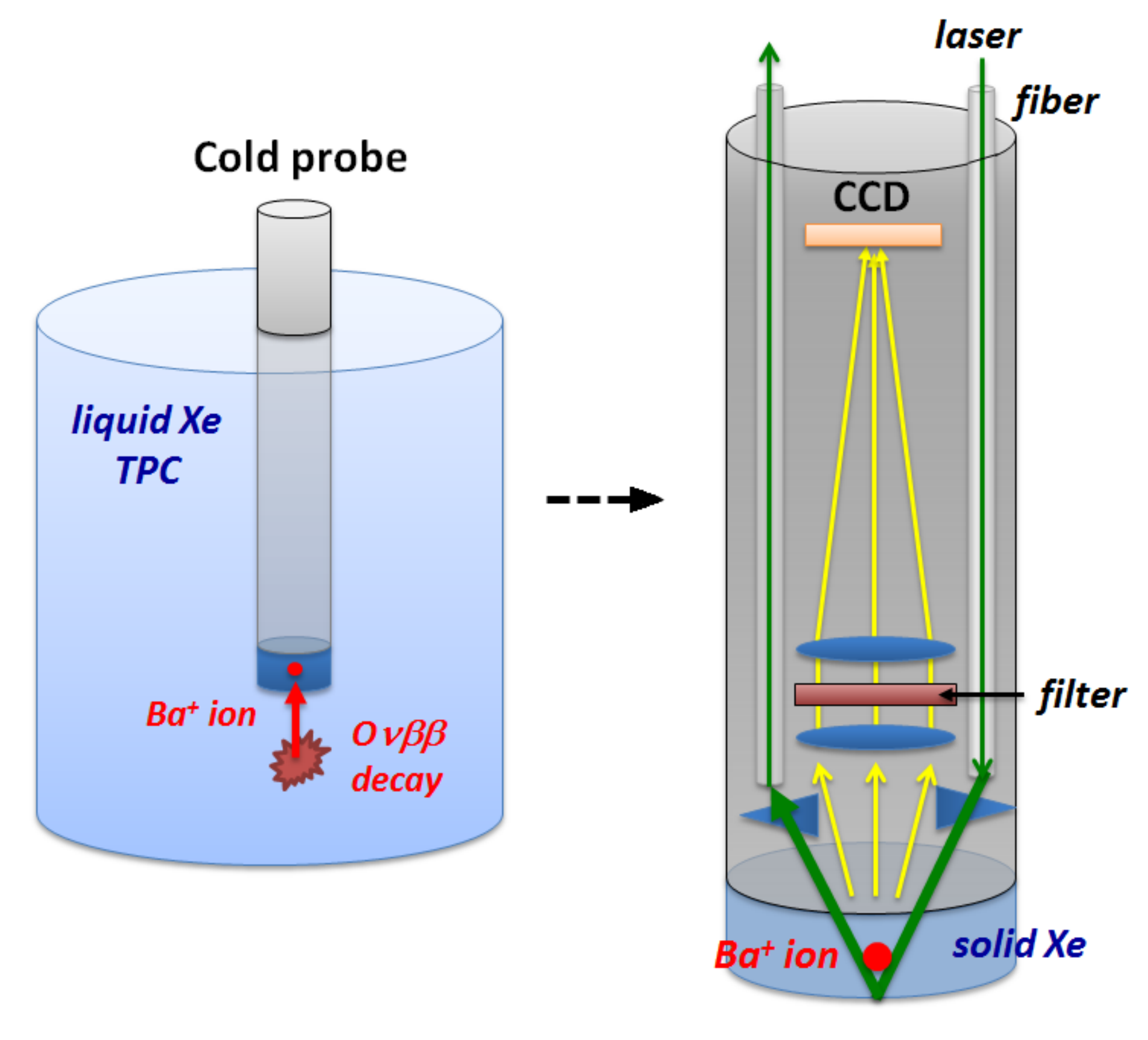}
	\caption{(color online) Left: concept of $^{136}$Ba daughter tagging on a cryoprobe in a liquid Xe double beta decay experiment.  Right: conceptual design of a cryoprobe for grabbing and detecting single $^{136}$Ba ions or atoms.}
	\label{fig:cryoprobe}
	\end{center}
\end{figure}

Several different methods of barium tagging in LXe are being explored by the nEXO Collaboration.
A method based on laser ablation and resonance ionization of Ba on a probe has recently been reported \cite{Twelker2014}.
In this paper progress is presented on a hybrid barium tagging method shown on the left in Fig.~\ref{fig:cryoprobe}.
In this method, the barium daughter ion from the $ 0\nu\beta\beta $ decay is frozen with some surrounding xenon on a cold probe that is inserted into the LXe.
This ion, or atom if further neutralization occurs, is then detected by matrix isolation spectroscopy in the solid xenon matrix on the probe.
The probe could also be extracted to a low pressure region where the Ba detection could be done at a lower temperature, at which fluorescence may be more efficient.

A concept for such a barium tagging probe is shown on the right in Fig.~\ref{fig:cryoprobe}.
When the probe is near the $^{136}$Ba$^+$ daughter ion, the flow of cooling gas, e.g., high pressure argon gas expanding through a Joule-Thompson nozzle or cryogenic helium gas, is increased to cool the end of the vacuum-insulated probe to below the Xe freezing point of 161 K.
This traps the Ba$^+$ ion (or Ba atom) in a thin layer of SXe on the sapphire window at the end of the probe.
The laser light needed to excite the single Ba$^+$ or Ba enters via an optical fiber within the probe.  
The light is deflected to illuminate the region of SXe where the Ba atom/ion is trapped.  
The thickness of the SXe layer is monitored by interference fringes in the reflected light guided by a second fiber.
Ba atom/ion fluorescence is collected by an efficient lens and focused on a CCD chip in the probe to produce an image of the single atom/ion.
The presence or absence of a peak in the CCD image distinguishes $^{136}$Xe double beta decay events from background events.

Previous measurements with Ba in matrix include absorption and emission spectra of the dominant visible transition of Ba in solid argon and krypton \cite{Balling1985} and  emission spectra of Ba$^+$ in solid and liquid helium \cite{Kanorsky1994,*Lebedev2011}. 
In \cite{Balling1985} it was reported that stable sites for Ba in solid xenon were not obtained, and that attempts to observe the emission spectrum failed. 

In this paper, detailed studies of the fluorescence of Ba atoms in SXe are reported.
Candidate Ba$^+$ lines are also found.  
Some results in solid argon (SAr) matrix are presented for comparison.
A series of molecular lines in Ba$^+$ deposits is related to hydrogen content in the matrix.
Images of small numbers of Ba atoms in solid xenon matrix are presented to demonstrate the progress towards single Ba atom or Ba$^+ $ ion tagging for a second phase of nEXO.

\section{Apparatus}
\begin{figure}[h!tb]
	\begin{center}
	\includegraphics[width=0.45\textwidth]{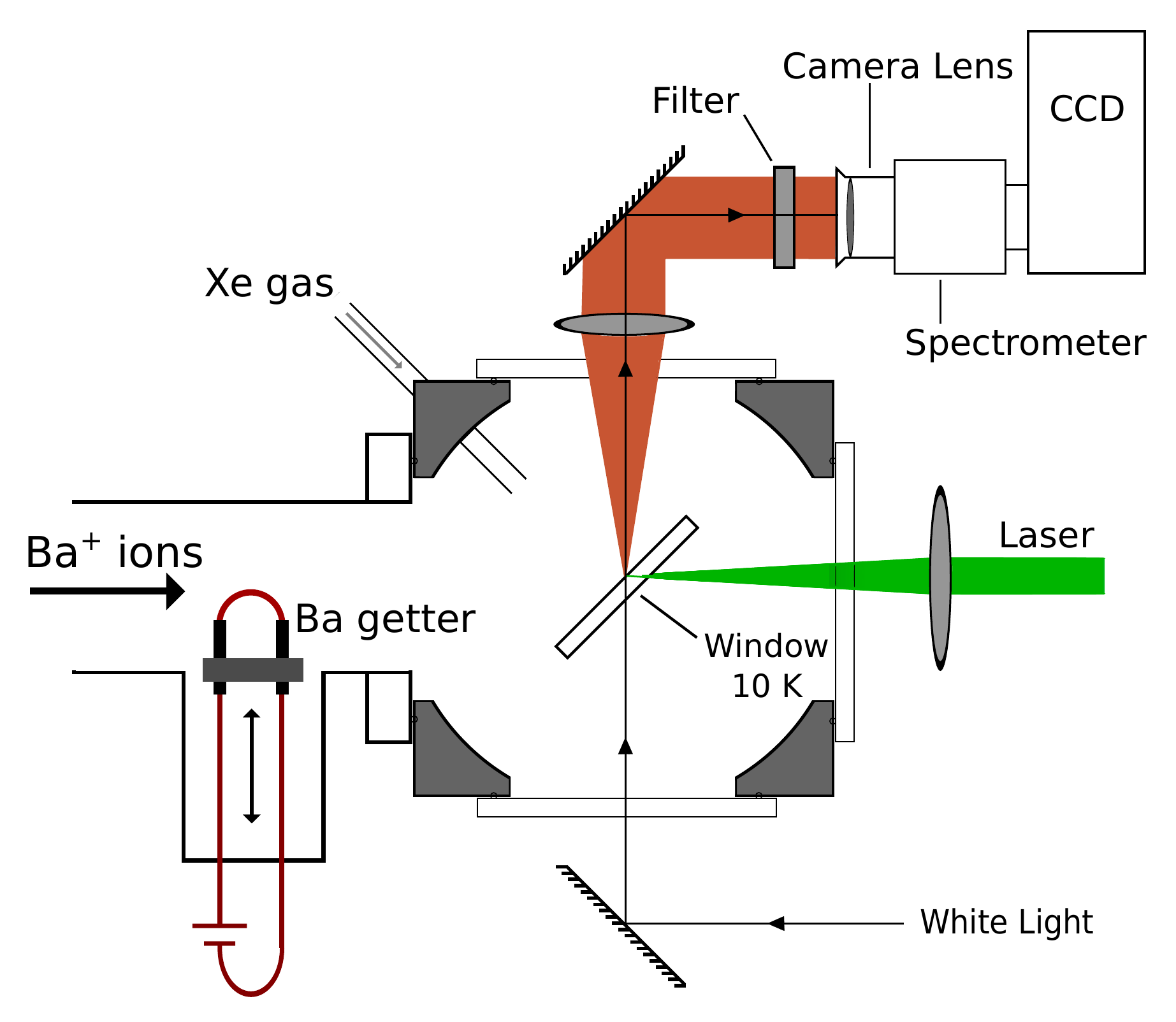}
	\caption{(color online) Apparatus used for spectroscopy of Ba and Ba$^+$ deposits in SXe.}
	\label{fig:optic_system1}
	\end{center}
\end{figure}

The apparatus for studying the spectroscopy and imaging of Ba atoms and Ba$^+ $ ions in SXe and SAr is shown schematically in Fig.~\ref{fig:optic_system1}.
A 19~mm diameter sapphire window, on which matrix samples were formed, was held in a copper housing (not shown) attached to the cold finger of a 10~K cryostat \cite{Cryostat}.
A small tube directed the gas used to form the matrix toward the center of the window.
For deposits of Ba atoms, a barium-aluminum getter source \cite{SAES} was heated by passing a DC current through it.
For Ba$^+ $ deposits, the getter assembly was pulled to the side, and Ba$^+$ ions from from a mass-selected ion beam were directed onto the matrix from the left.
The sapphire window was angled at 45$^{\circ}$ such that its front surface had line of sight to the barium sources, while still being visible to the collection optics and gas tube.  

The cold finger was equipped with a band heater so that the temperature of the window, measured by a thermocouple, could be stabilized at temperatures other than 10~K, e.g., for deposits or annealing at higher temperature. 
A silicon diode was also installed on the window holder for more accurate temperature measurements ($\pm 0.5$ K).
The window was surrounded by a thermal shield (not shown) and contained in a 2 inch evacuated cube with five viewports.

The laser beam used for exciting the emission of atoms or ions entered through the port on the right.
Sometimes a 7~cm lens was used to focus the laser beam to smaller diameters in the matrix.
A 5~cm focal length f/2 lens at the top collected the fluorescence into a collimated beam that was directed by two steering mirrors into a 50~mm focal length Nikon camera lens or a Phoenix 70-210~mm zoom lens.
The light was focused at the input slit of an imaging spectrometer of 150~mm focal length with a liquid nitrogen cooled CCD detector at the output plane \cite{Acton}. 
Raman and bandpass filters were placed between the lenses to block scattered laser light and other undesired light, such as the 693~nm fluorescence line of a very low concentration of Cr$^{3+} $ impurities in the sapphire window.
The fluorescence detection efficiency in this configuration was $0.35-1.4\times10^{-3}$ counts/photon emitted, depending on camera lens and zoom used, with an uncertainty of a factor of two.
This included factors for the collection solid angle of the lens, narrowed collection angle in the SXe due to refraction at the SXe/vacuum interface, the transmission of filters and optics, the transmission of the spectrometer, the quantum efficiency of the CCD chip and the CCD digitization factor in photoelectrons/count. 

Various CW laser sources were used for excitation, including seven discrete argon ion laser lines from 454~nm to ~514~nm, a doubled Nd:YAG laser at 532~nm, a tunable Coherent 599 dye laser pumped by the argon laser for 545-564~nm, and a second tunable Coherent 599 dye laser pumped by a krypton ion laser for 459-514~nm.
Typical laser powers were a few tens of milliwatts, sometimes attenuated by neutral density filters up to OD4.  
Laser powers were measured with a power meter via a calibrated pickoff window and recorded simultaneously with fluorescence spectra using a LabView program.
The laser beam 1/e$^2$ radius, $w$, was 1-4~mm for the unfocused laser beams and a few $\mu$m at the focus when the lens was used.
Intermediate laser radii were obtained by moving the lens.
The LabView program controlled the delay of a shutter in the laser beam path relative to the shutter and frame period of the CCD camera.
This allowed precise timing of the laser exposure so that fluorescence from the very beginning of the optical pumping process could be captured.

The absorption measurements presented in this work were done with a different optics system in which white light entered from the bottom port.
The transmitted white light was coupled by a fiber coupler at the top port into a multi-strand, multi-mode fiber bundle.
At the other end of the bundle, the fibers formed a vertical line image at the entrance plane of the spectrometer.
The white light source for absorption measurements was a halogen lamp that was collimated by an aperture to a small solid angle and filtered to provide a flatter spectrum in the anticipated absorption range (450~nm-650~nm).
One set of emission spectra reported in this paper and the Ba in SAr emission spectrum were also taken in this configuration with the fiber coupler in an off-axis position at the top port.

\begin{figure*}[t]
	\begin{center}
	\includegraphics[width=1\textwidth]{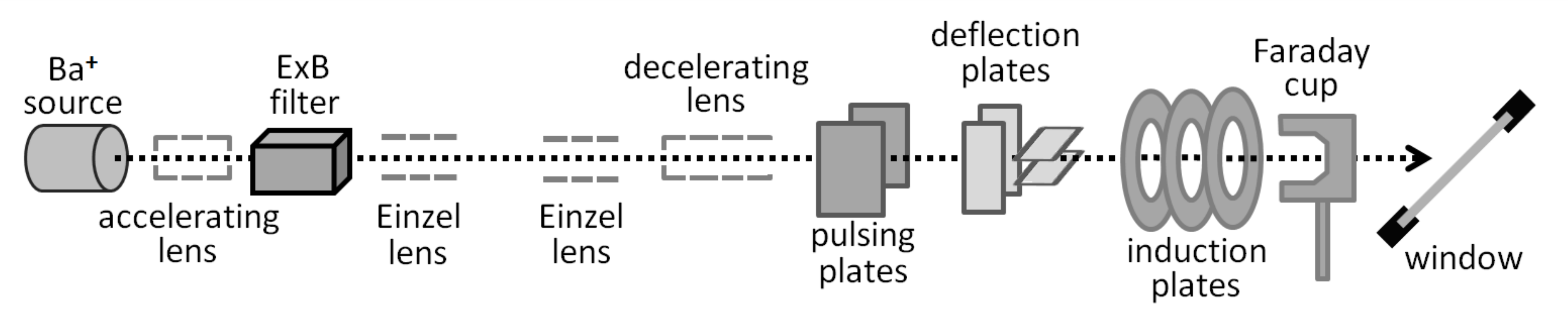}
	\caption{Ion beam apparatus}
	\label{fig:ion_beam_system}
	\end{center}
\end{figure*}

The ion beam system for depositing Ba$^+ $ ions in SAr or SXe is shown in Fig.~\ref{fig:ion_beam_system}.
The ion source, extraction optics, lens and E$\times$B mass filter were from a commercial Colutron ion gun \cite{Colutron}.
Additional Einzel lenses, deflection plates and a deceleration lens provided beam steering, focusing and deceleration so that Ba$^+ $ ions could be deposited in the matrix at energies from 100~eV to 2~keV.
The data on Ba$^+$ deposits presented in this paper were taken with 2~keV ion beam energy,
The beam current was monitored with a Faraday cup of 2~mm diameter.
Typical beam currents were on the order of 10~nA.

To achieve deposits of small numbers of ions, a pair of pulsing plates were set at $\pm$200~V DC to deflect the beam and then pulsed to 0~V for 1 $\mu$s to pass a short burst of Ba$^+ $ ions.
The pulse rate was controlled by a LabView program and could be operated from single pulse to 500~Hz in burst mode or continuous mode.
The induction signal on a central conductor shielded by two grounded plates was recorded on a digital oscilloscope and provided a monitor of the ion pulses during the experiment.

At times the sapphire window was replaced by a Faraday cup at the window position to check the beam location and size at the sample relative to the beam Faraday cup both for DC and pulsed beams.
Scintillation light from the ion beam hitting the SXe matrix was also used to confirm optimum ion beam deflection to the center of the window area imaged by the optical system.

Vacuum was maintained with a small turbomolecular pump attached to the cryostat and two large turbomolecular pumps on the ion beam system.  Typical residual gas pressures were around $ 1 \! \times \! 10^{-7} $ Torr with the warm cryostat. 
The gas supply was research grade (99.995\% pure) xenon or argon.
Gas flow was controlled by a leak valve \cite{Granville-Phillips}.
No purifier was used because the gas flow rate was too low for effective purification. 

\section{Method}
\begin{figure}[h!tb]
	\begin{center}
	\includegraphics[width=0.45\textwidth]{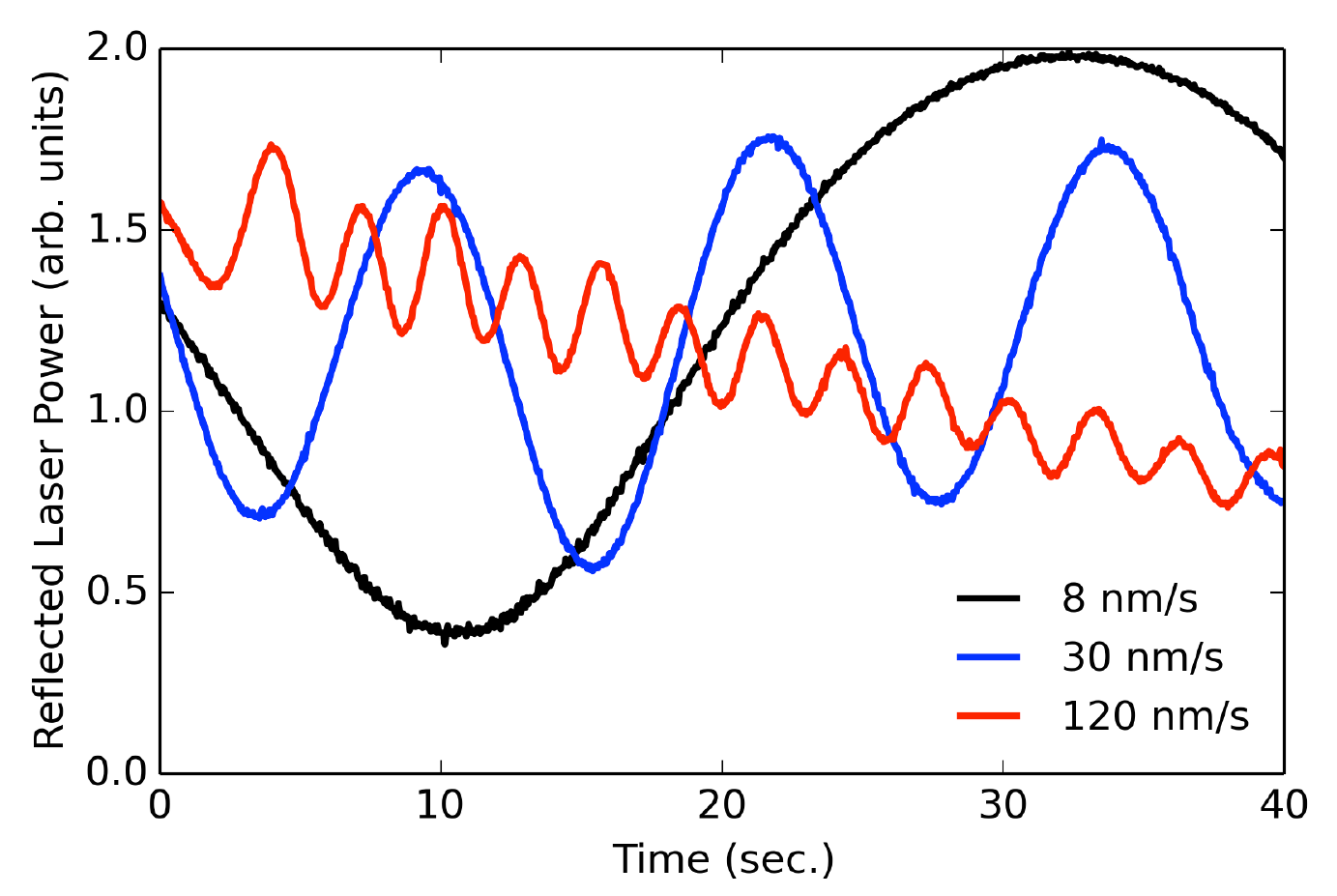}
	\caption{(color online) Interference fringes with a  532~nm laser and resulting matrix growth rates at different Xe leak rates.}
	\label{fig:fringe_rates}
	\end{center}
\end{figure}

Solid rare-gas matrices were prepared by leaking gas (argon or xenon) onto the cold sample window. 
The matrix growth rate was finely controlled by adjustment of the leak valve.  
Interference fringes in transmission or reflection of a laser beam were used to measure the matrix growth rate. 
Sample fringes at three leak settings corresponding to matrix growth rates of $\sim $ 8-120~nm/s are shown in Fig.~\ref{fig:fringe_rates}.
The fringe rate for residual gases only was observed to be much less than one fringe in two hours, perhaps on the order of 0.005~nm/s.
Using the ratio of these numbers, the residual gas impurity content in the matrix was $\le$100-1000~ppm at the above leak rates.

The gas flow was turned on for 5 seconds to 5 minutes, depending on the leak rate, to build a pure host matrix foundation prior to introducing barium. 
For atomic deposits, Ba was then co-deposited with the gas by joule-heating a length of getter using up to 15 amps DC.  
Pre-heating of the getter at lower currents was done before any deposit was made to bake out adsorbed gases, particularly hydrogen.
For ion deposits, the Faraday cup was used as a shutter to control the deposition time.
Large deposits, where absorption could be detected, typically lasted for a few minutes.
The Faraday cup was pulled out for $\le$ 1~s for smaller deposits.
The matrix was then capped with a final layer of pure noble gas. 

\section{Results}

\begin{figure}[h!tb]
	\includegraphics[width=0.45\textwidth]{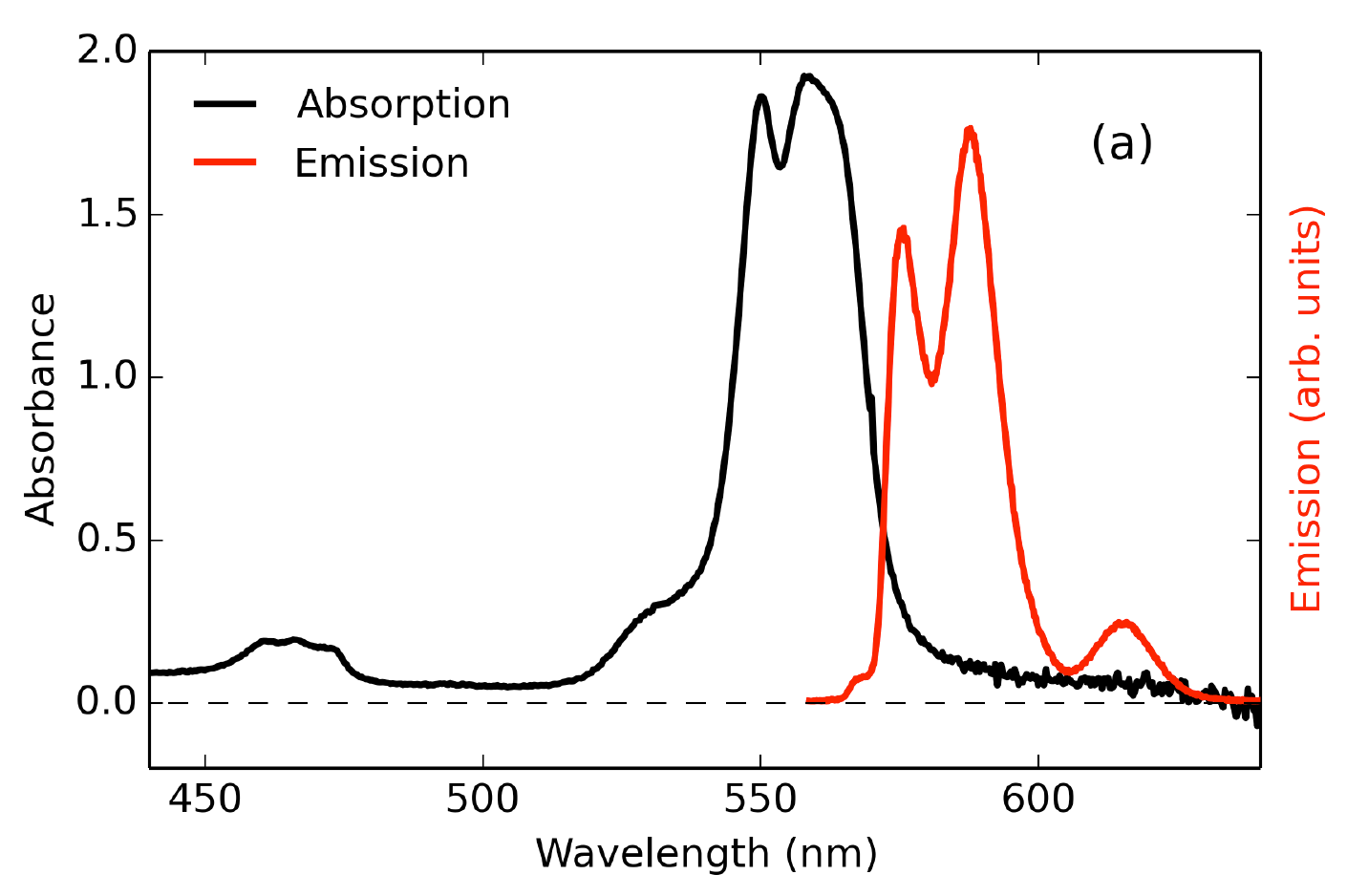}
	\includegraphics[width=0.45\textwidth]{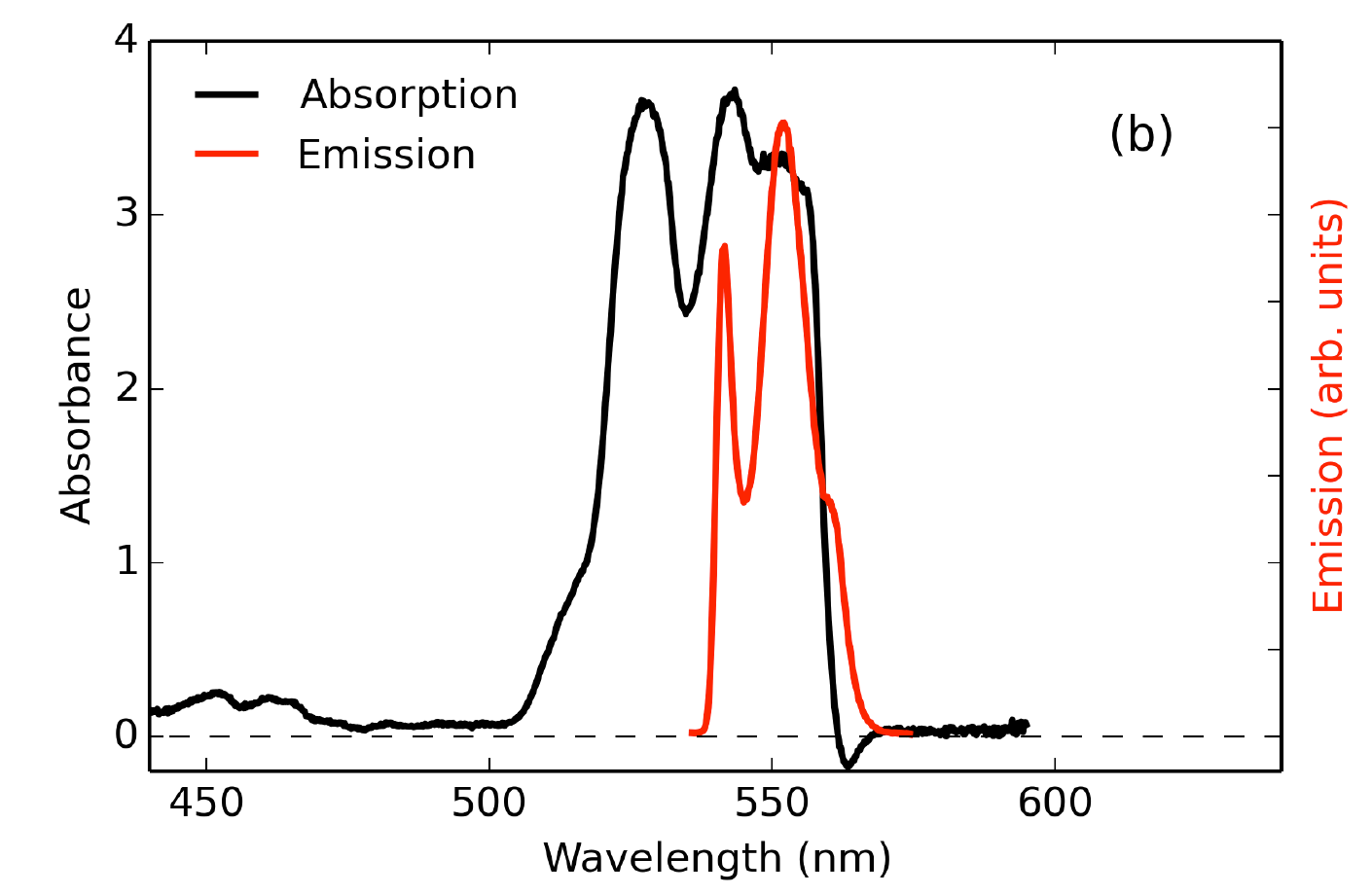}
	\caption{(color online) Absorption (black) and representative emission (red) spectra of barium in solid (a) xenon and (b) argon.  The absorbance $A_\lambda=-ln(T_\lambda)$, where $T_\lambda$ is the transmission at wavelength $\lambda$.}
	\label{Fig:BaAbsFlur}
\end{figure}

Absorption spectra of Ba in SXe and SAr from a getter deposit at 10~K are shown in Fig.~\ref{Fig:BaAbsFlur}. 
Multiple components in the dominant band are visible in both matrices.  
Qualitatively similar absorption was found in SXe with a Ba$^+$ ion deposit, suggesting substantial neutralization upon entering the matrix.
The absorption band in SAr has been identified with the  $6s^2 ~{^1}S_0 \rightarrow 6s6p ~{^1}P_1$ transition in atomic Ba, that occurs at 553.5~nm in vacuum \cite{Balling1985}.
A similar assignment for SXe is reasonable.
In xenon, the central peak has a 4~nm red shift from the vacuum value, whereas in argon the central peak has an 11~nm blue shift.  
A smaller blue absorption band with three peaks is also seen in both spectra.  

Representative emission spectra of barium in SXe and SAr at 10~K are shown in Fig.~\ref{Fig:BaAbsFlur}. 
The excitation laser wavelengths for these spectra are 555~nm and 532~nm, respectively.
It can be seen that the Stokes shift, the red shift of emission relative to absorption, is greater in the heavier noble gas, as is typically the case in matrix isolated atomic spectra \cite{Crepin1999}. 
Several partially resolved emission peaks are evident.

These emission spectra are for Ba$^+$ ion and Ba getter deposits in SXe and SAr made at 45~K and 10~K, respectively, and observed at 10K before substantial bleaching occurs. 
As discussed below, the relative strength of different peaks in the emission spectra can vary substantially depending on excitation wavelength, deposition temperature, measurement temperature, history of annealing and bleaching.  

\begin{figure}[h!tb]
	\begin{center}
	\includegraphics[width=0.45\textwidth]{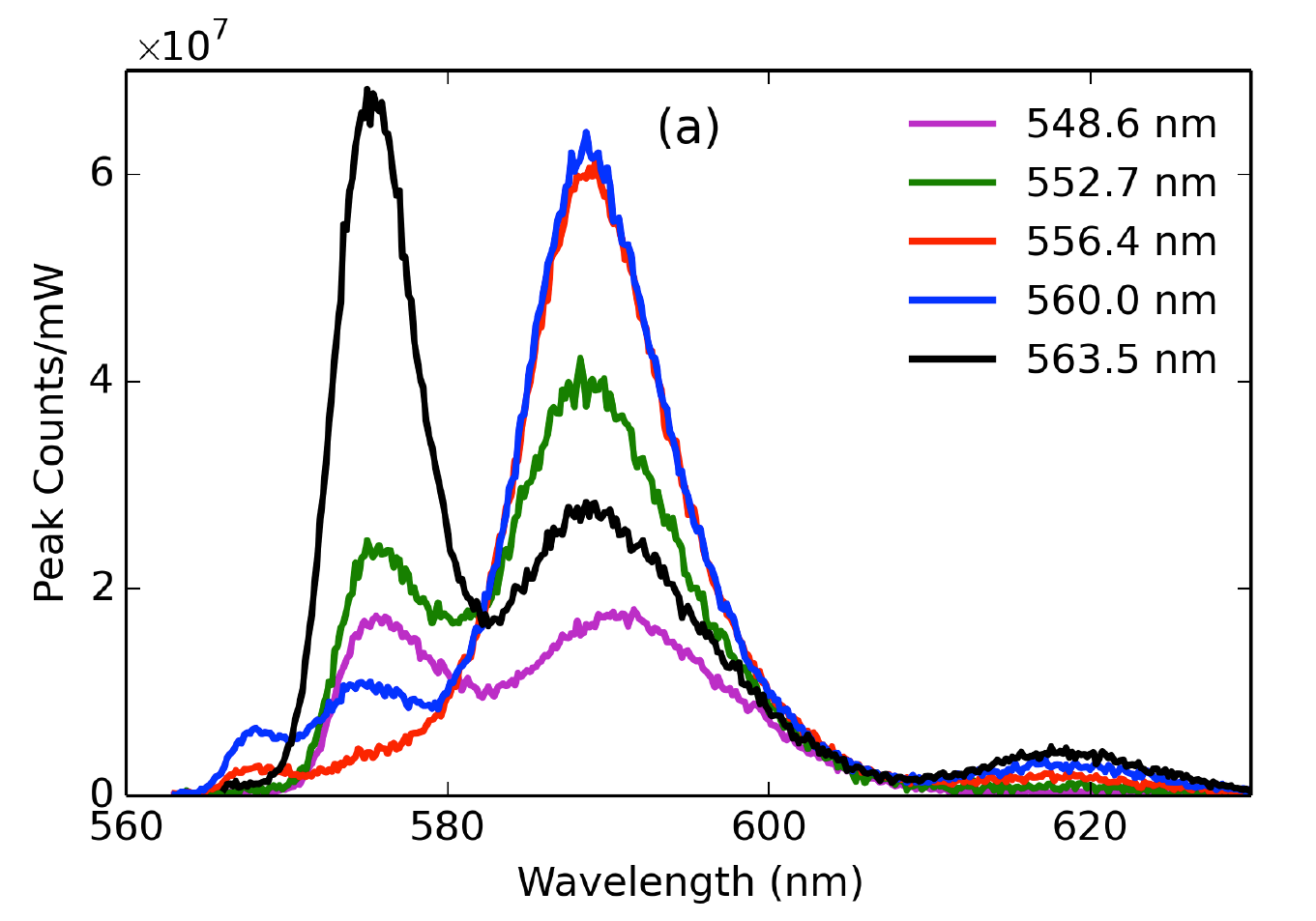}
	\includegraphics[width=0.45\textwidth]{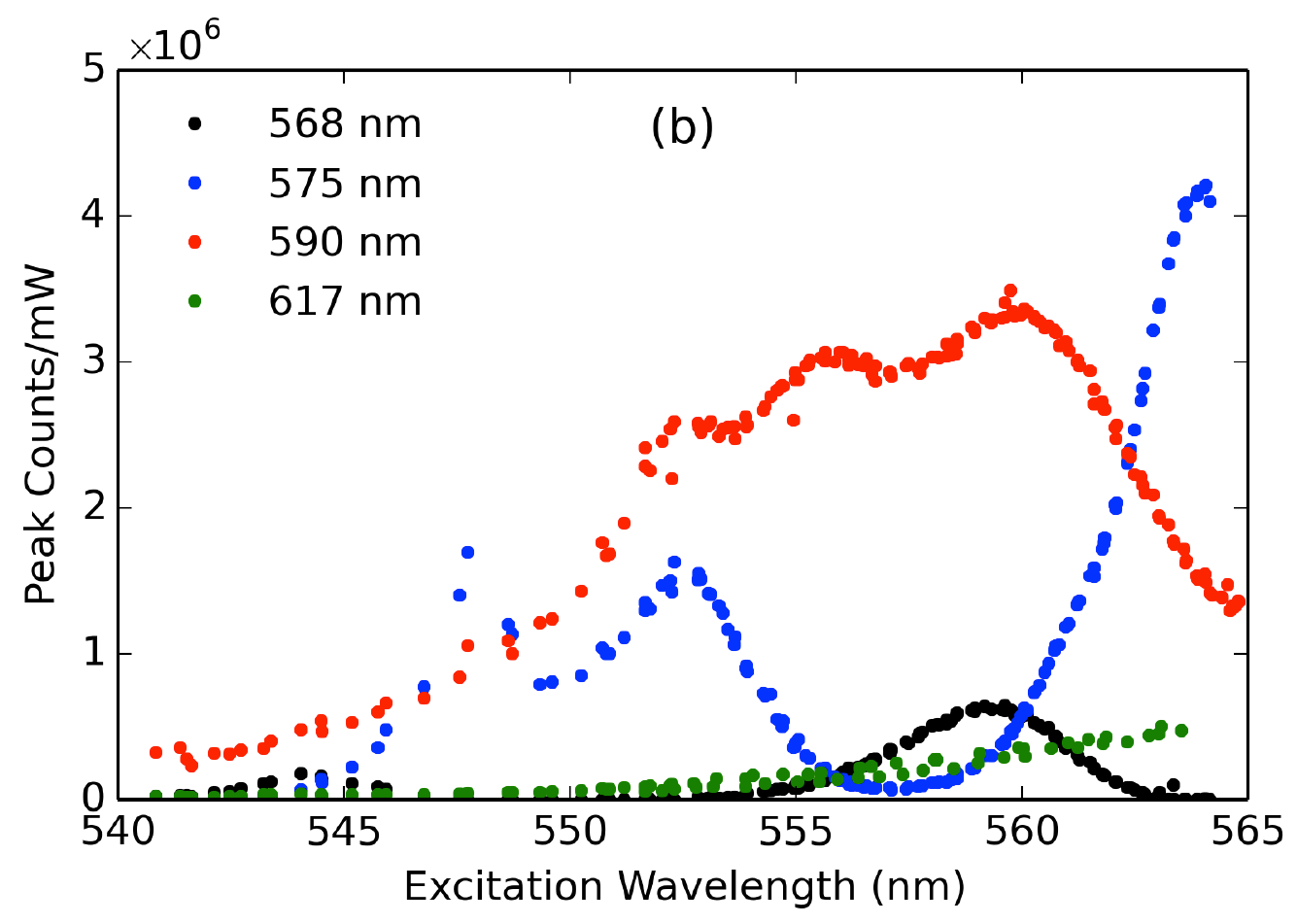}
	\caption{(color online) (a) Emission spectra of a Ba$^+$ ion deposit in SXe made at 44~K and observed at 10~K for a sample of excitation wavelengths; (b) excitation spectra for the 568~nm, 575~nm, 590~nm  and 617~nm peaks of Ba in SXe.}
	\label{fig:SXeExcitation}
	\end{center}
\end{figure}

Ba emission spectra in SXe for five sample excitation wavelengths are shown in Fig.~\ref{fig:SXeExcitation}(a).
Large changes in the emission spectrum are apparent.
Excitation spectra, i.e., the peak counts/mW of the prominent emission peaks for all the laser wavelengths in the run, are shown in Fig.~\ref{fig:SXeExcitation}(b).  
The three emission peaks have quite different excitation spectra.
This supports associating each emission peak with Ba atoms in a different site, or neighbor atom configuration, in the SXe matrix.

\begin{figure}[h!tb]
	\includegraphics[width=0.45\textwidth]{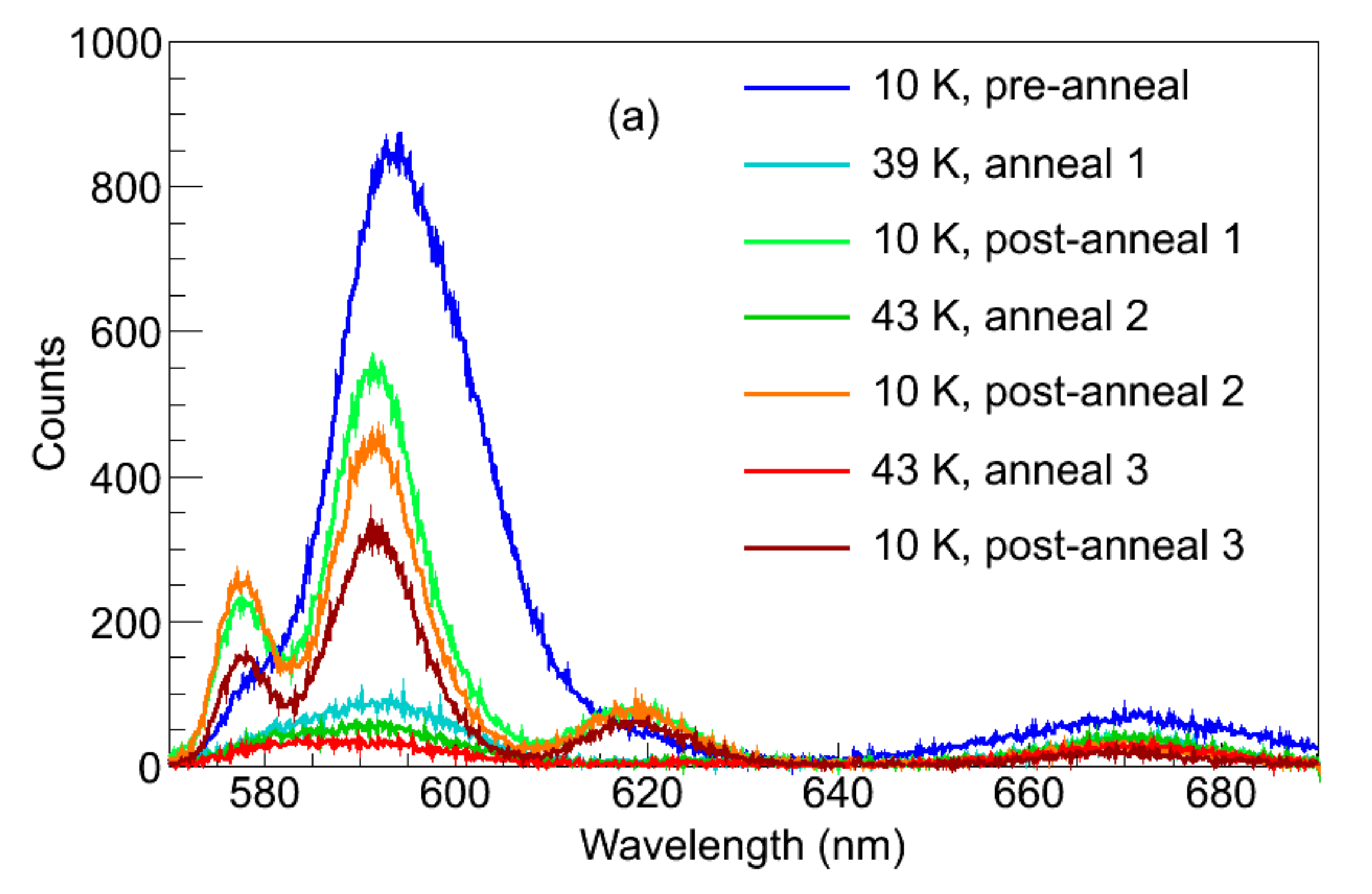}
	\includegraphics[width=0.45\textwidth]{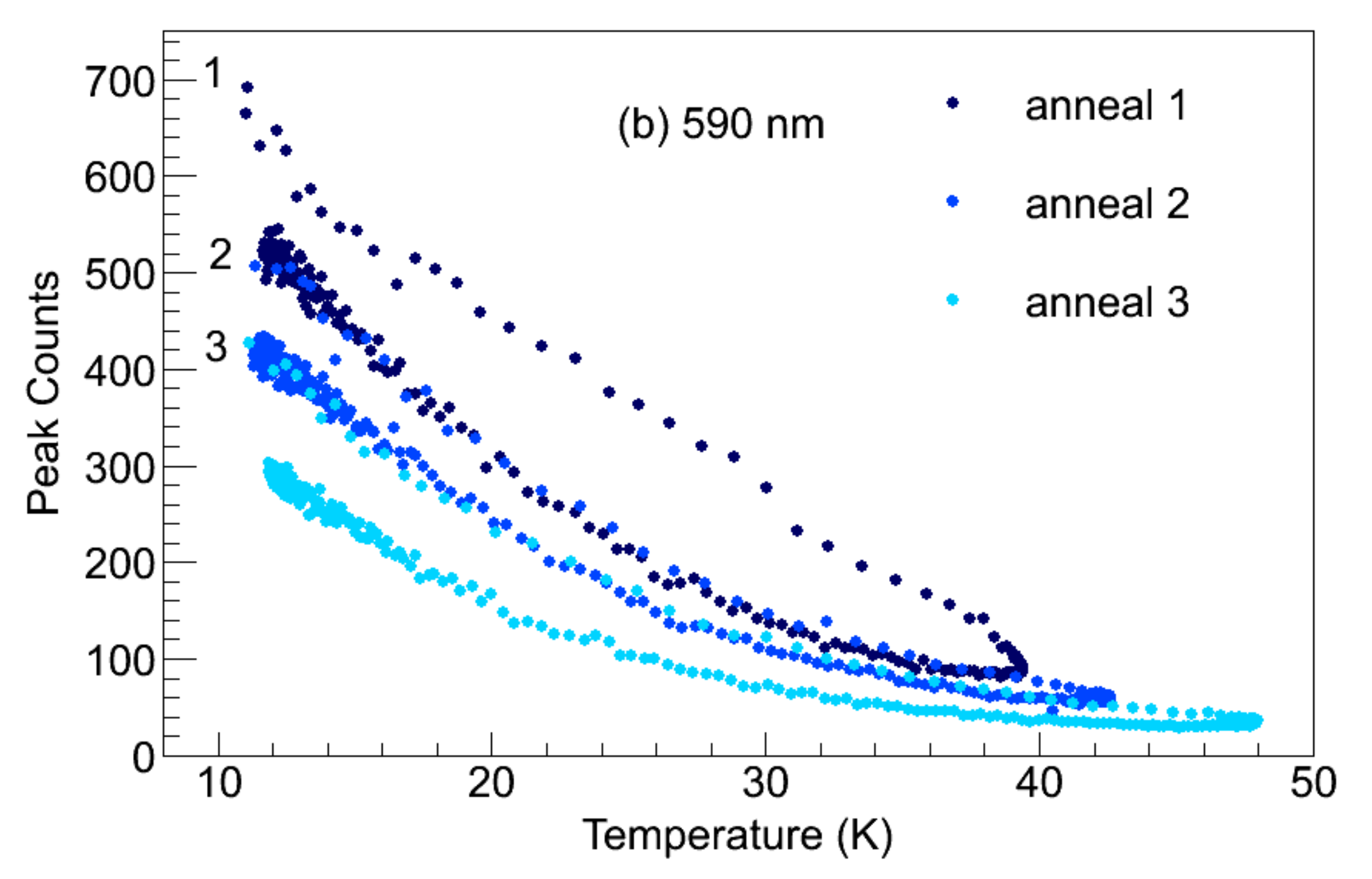}
	\includegraphics[width=0.45\textwidth]{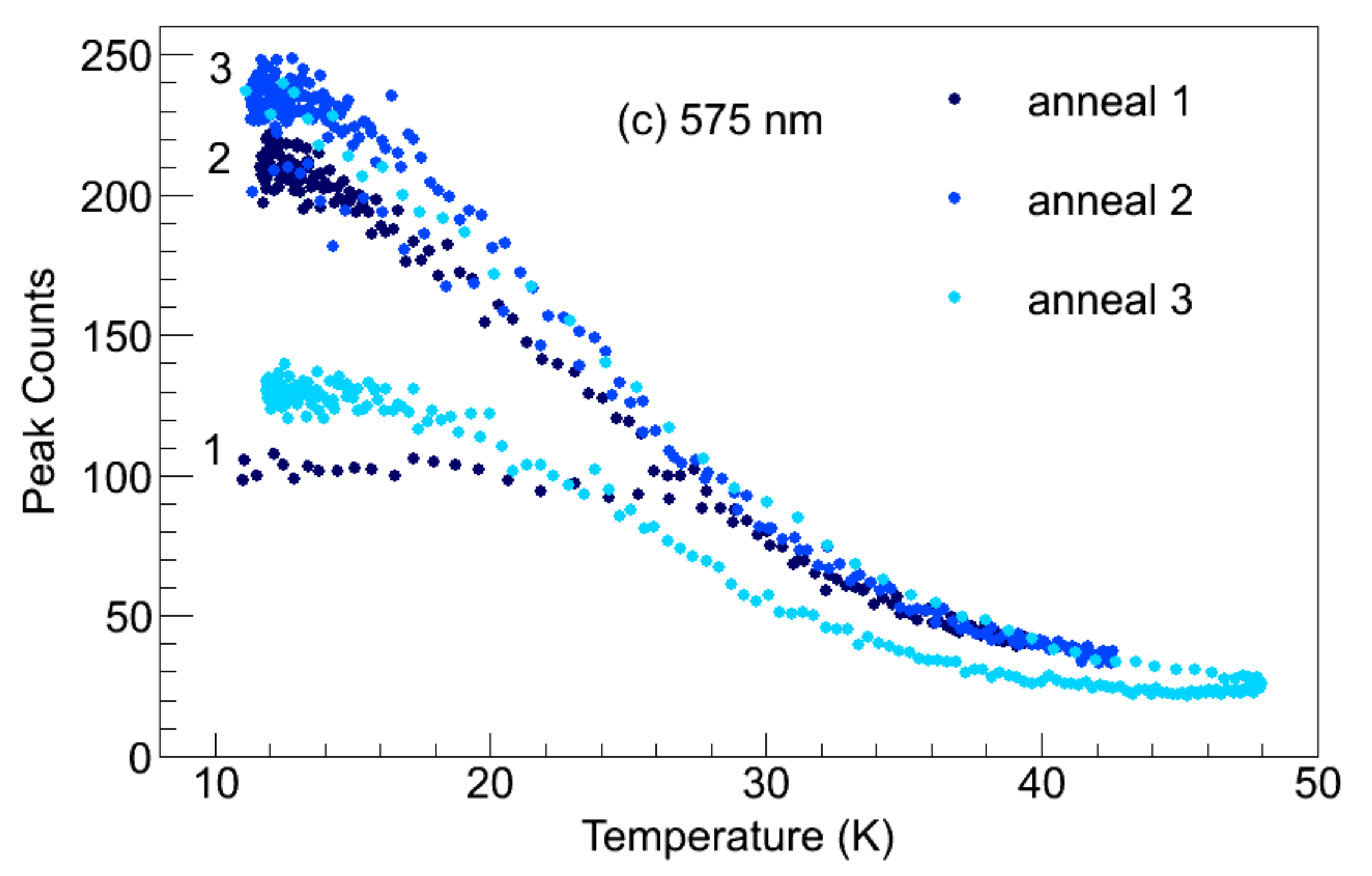}
	\caption{(color online) Annealing of a Ba$^+$ in SXe deposit made at 11~K: (a) spectra at the lowest and highest temperatures for 3 anneal cycles; peak counts in Gaussian fits to spectra during the three anneal cycles for the (b) 590~nm and (c) 575~nm peaks.  The excitation wavelength was 564~nm.}  
	\label{Fig:SXeAnnealCounts}
\end{figure}

Additional information on the origin of the different emission peaks can be gained by studies of annealing and through deposits made at varying temperature. 
The effect of annealing on the emission spectrum of a Ba$^+$ deposit in SXe is shown in Fig.~\ref{Fig:SXeAnnealCounts}(a). 
The peak counts of the 590~nm and 575~nm components a four-Gaussian fit to the spectra during the annealing cycles are shown in Figs.~\ref{Fig:SXeAnnealCounts}(b) and (c), respectively.
The spectrum, as deposited, had a large and broad 590~nm emission peak with a tail extending to the red and a hint of a 575~nm peak.
Upon warming to 39~K, the spectrum was reduced to a single broad bump.
After cooling back to 10~K, the spectrum had changed considerably.
The 590~nm peak was narrower and lower, and the 575~nm peak was increased to twice its initial size.
Two more annealing cycles to 43~K and 48~K yielded a modest further reduction of the 590~nm peak and growth of the 575~nm peak.
It is noteworthy that for both wavelengths, the warming portion of the second cycle closely followed the cooling part of the first cycle, and similarly in the next cycle.
This confirms that bleaching was negligible in these experiments with low laser power and unfocused laser beam. 
Thus the observed spectral modifications in annealing cycles can be attributed to changes in the matrix sites of the Ba atoms.

\begin{figure}[h!tb]
	\includegraphics[width=0.45\textwidth]{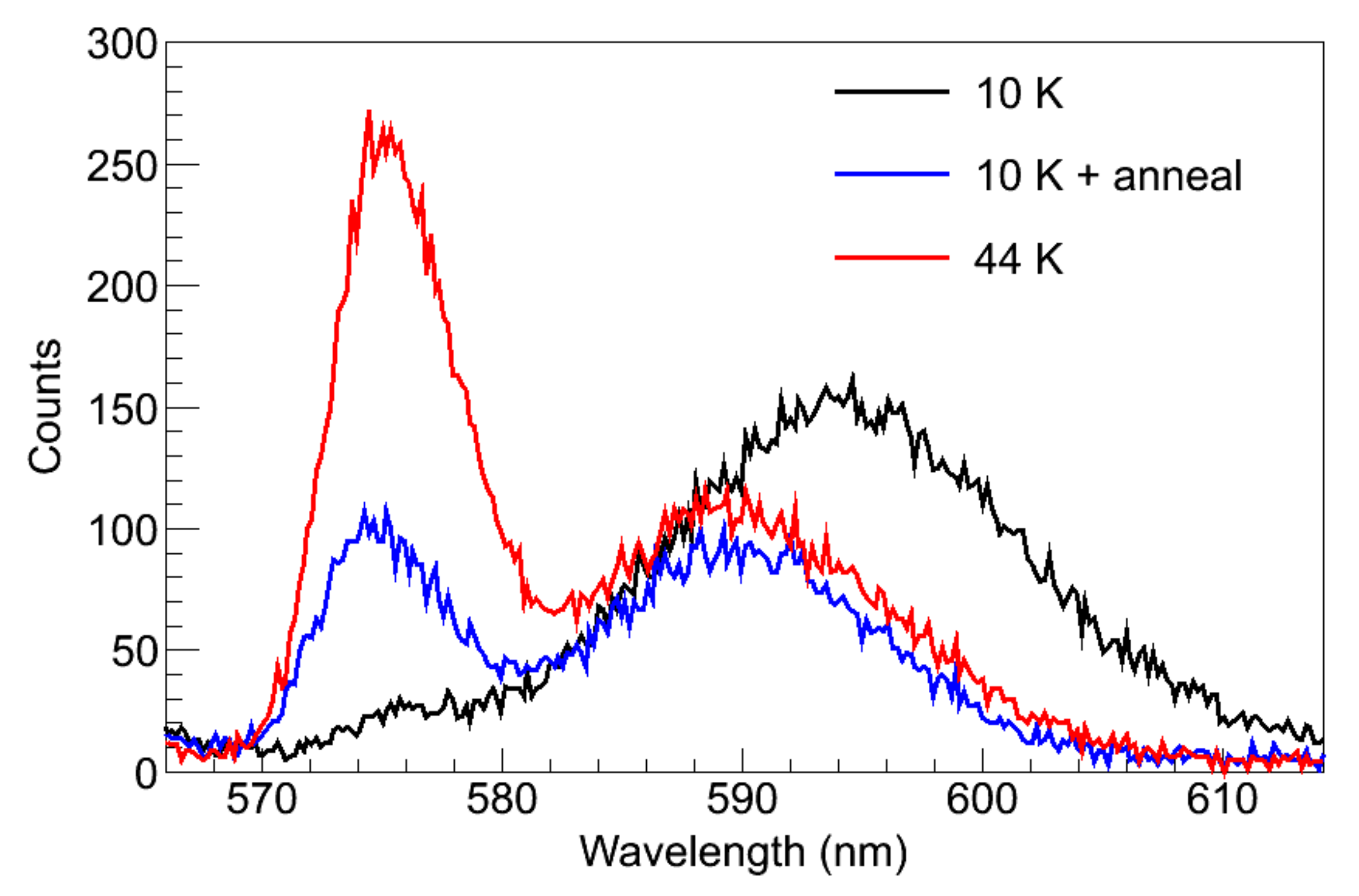}
	\caption{(color online) 10~K emission spectra of Ba$^+$ deposits made at 10~K (black) and 44~K (red), and the spectrum of the 10~K deposit after annealing to 39~K (blue).  All spectra were taken at 10~K.  The excitation wavelength was 564~nm.}
	\label{Fig:SXe_10K_44K}
\end{figure}

The temperature of deposition affects the relative abundance of Ba atoms in different matrix sites.
Emission spectra of a Ba$^+$ deposit made at 44~K and then observed at 10~K and a deposit made and observed at 10~K are shown in Fig.~\ref{Fig:SXe_10K_44K}.
The 44~K deposit has a much larger 575~nm peak, whereas the 10~K deposit has a broader and more red-shifted 590~nm peak.
Annealing of the 10~K deposit to 39~K and back yields a  spectrum more like that of the 44~K deposit.
The red tail of the 590~nm peak in Figures~\ref{Fig:SXeAnnealCounts}(a) and \ref{Fig:SXe_10K_44K}  may indicate the existence of an additional matrix site with an emission peak at greater than 590~nm that is populated in the 10~K deposit but is depopulated by annealing.

\begin{figure}[h!tb]
	\begin{center}
	\includegraphics[width=0.45\textwidth]{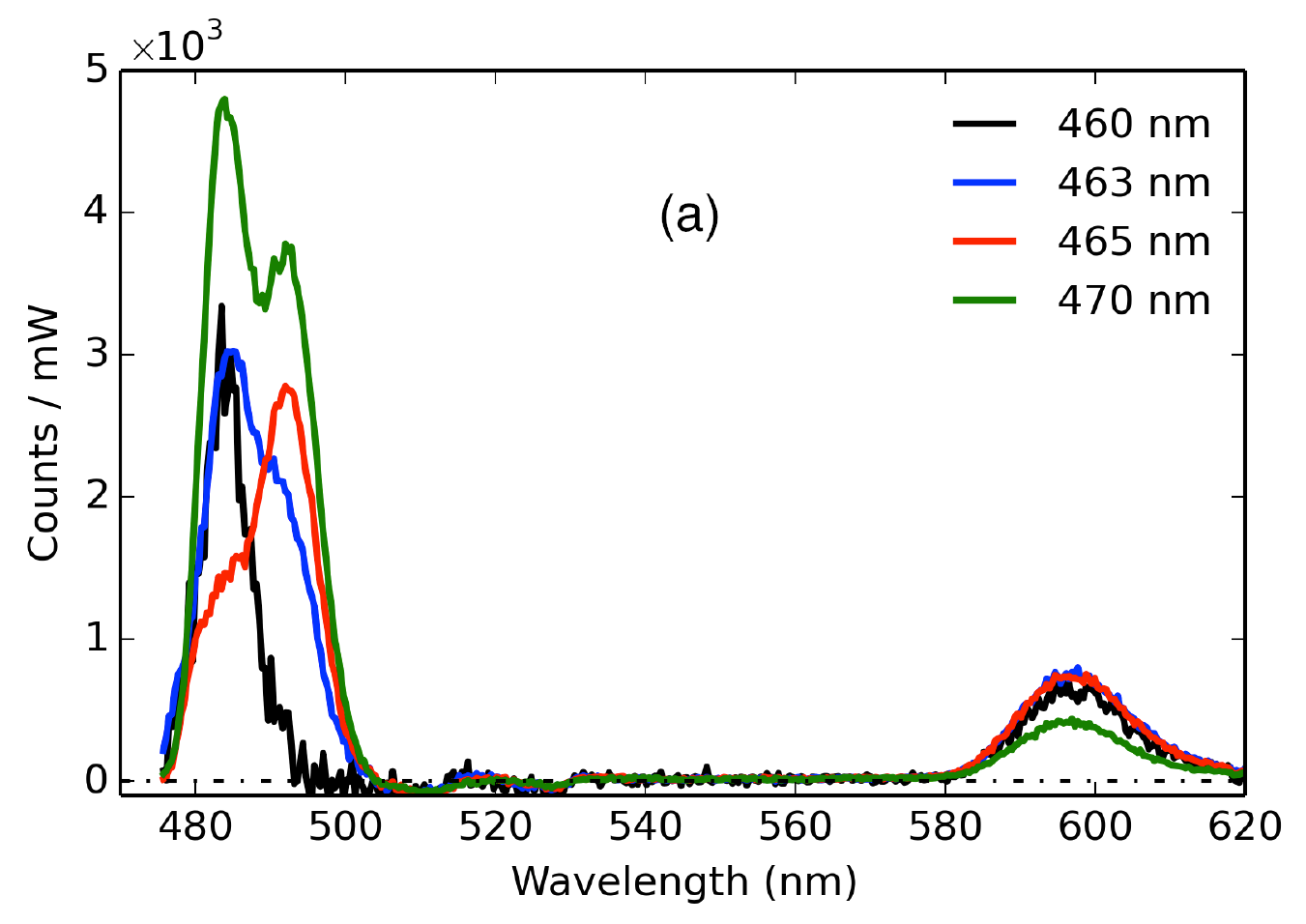}
	\includegraphics[width=0.45\textwidth]{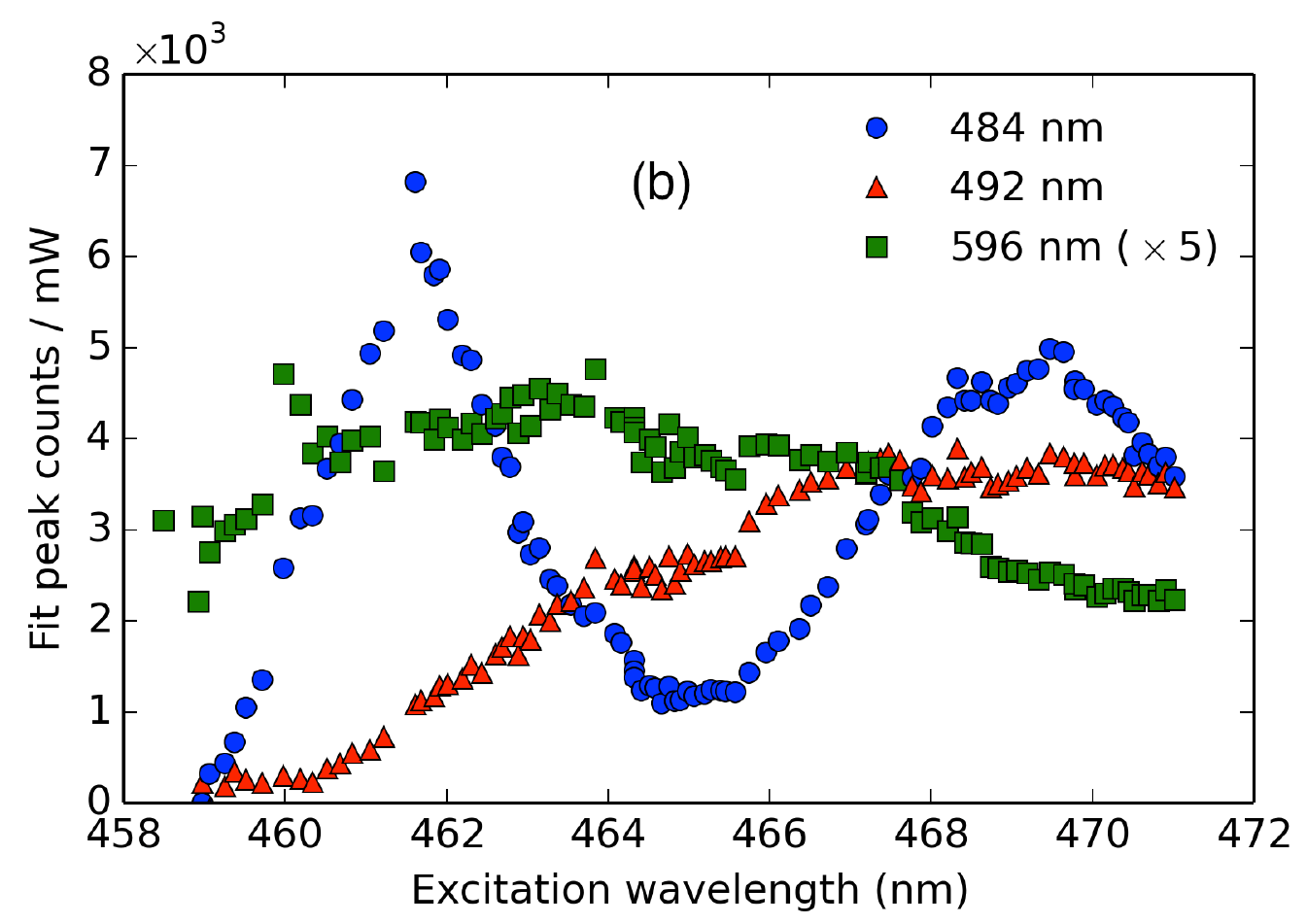}
	\caption{(color online) (a) Sample emission spectra of a 10~K Ba getter deposit in SXe excited by a blue dye laser; (b) Excitation spectra of the candidate Ba emission lines at 484~nm and 492~nm and the 596~nm Ba line based on Gaussian fits to the complete set of spectra.  These spectra were taken with the fiber coupling system of optics.}
	\label{fig:blue_lines}
	\end{center}
\end{figure}

Additional barium fluorescence peaks in both SXe and SAr have been observed using lower wavelength excitation from an argon ion laser and a blue tunable dye laser.
Sample emission spectra for a Ba getter deposit in SXe, excited with a tunable blue dye laser, are shown in Fig.~\ref{fig:blue_lines} (a).
New emission peaks at 483~nm and 492~nm are observed.
The emission peak at 596~nm may be the same as the primary Ba resonance line discussed above, only weakly excited by blue wavelengths.
For higher excitation wavelengths, a Raman filter with a higher cutoff wavelength (514~nm) was used.
This filter blocked the two blue emission lines.
No new emission peaks were observed with excitation up to 514~nm.

Excitation spectra for the two blue lines and the 596~nm line extracted from the full set of emission spectra are shown in Fig.~\ref{fig:blue_lines}(b).
The differing excitation spectra for the blue lines may imply separate atomic transitions or different matrix sites for a single transition.
It is interesting that the excitation spectrum of the 596~nm line matches quite well with the weak blue absorption band in Fig.~\ref{Fig:BaAbsFlur}(a). 

Comparison of deposits made with the Ba getter source and the Ba$^+ $ ion beam source may help distinguish absorptions and emissions due to Ba and  Ba$^+ $.
The getter source should produce mainly Ba atoms, and the energy of the deposit should be thermal (less than 1eV).
As discussed above, it is observed that a Ba$^+ $ ion deposit in SXe leads to neutral Ba fluorescence around 590~nm.
Thus some neutralization does occur.
Nevertheless, additional lines in the ion deposit not seen in the neutral deposit would be candidates for assignment to Ba$^+ $.

\begin{figure}[h!tb]
	\includegraphics[width=0.45\textwidth]{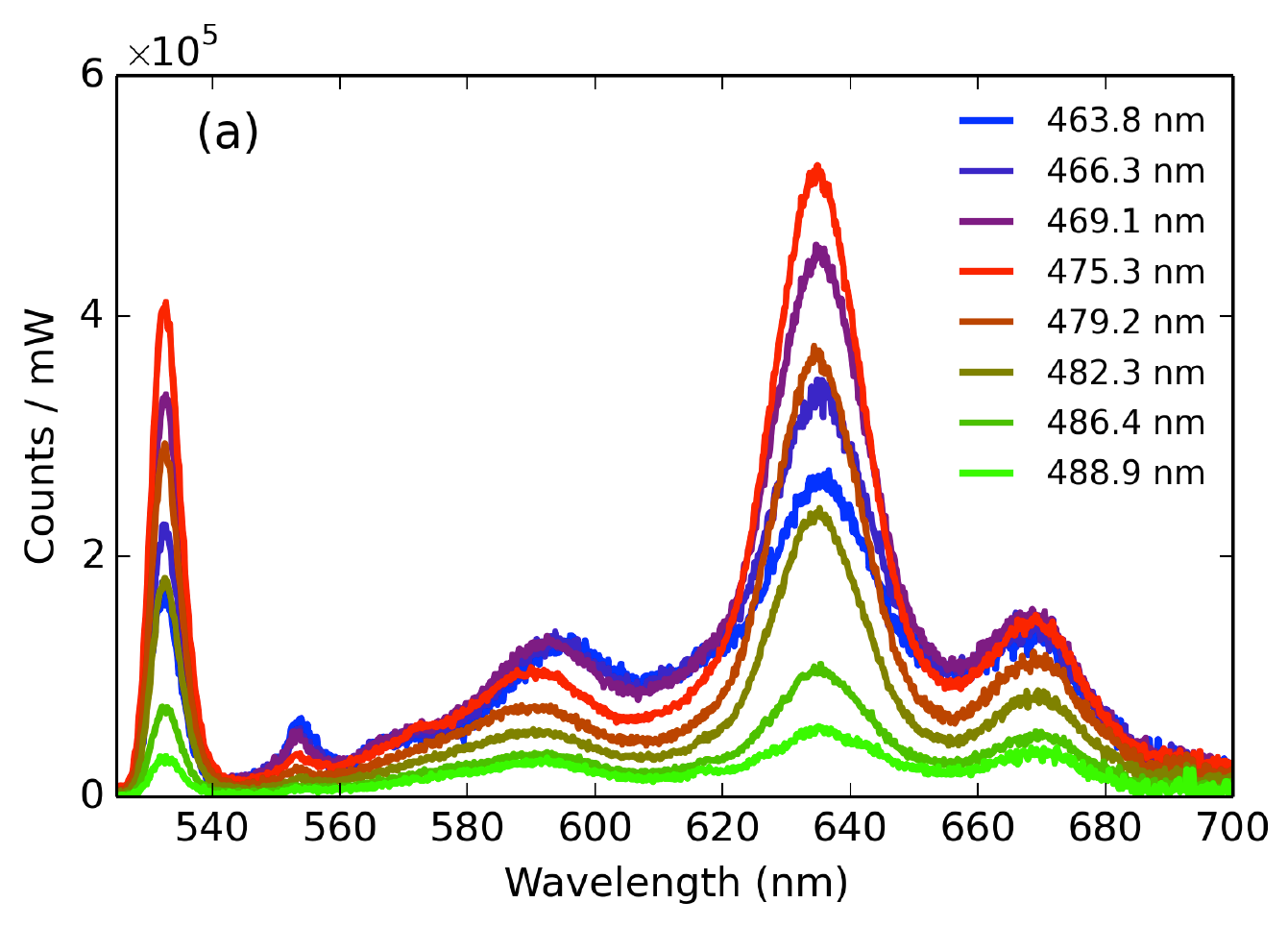}
	\includegraphics[width=0.45\textwidth]{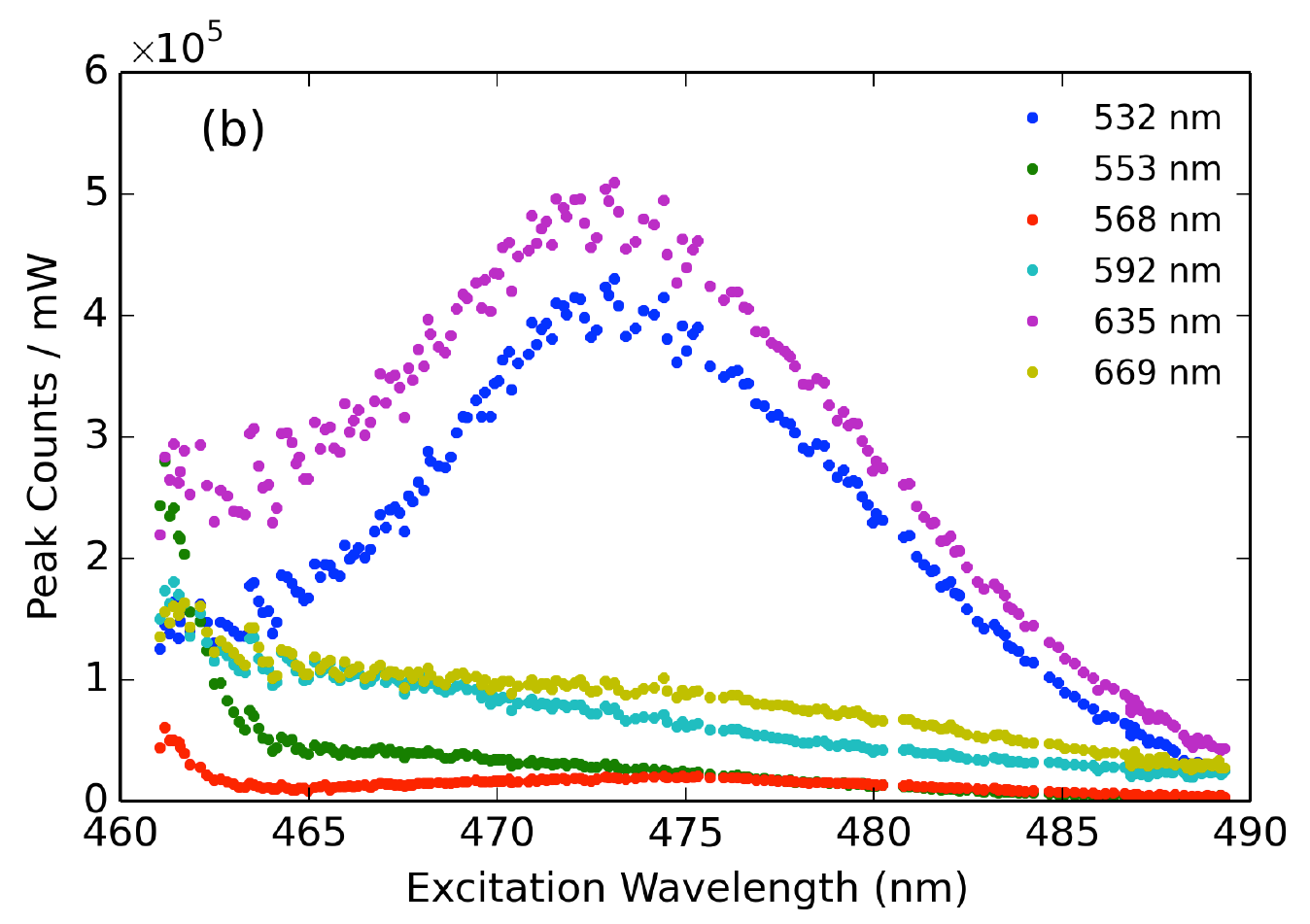}
	\caption{(color online) (a) Sample emission spectra of a Ba$^+ $ ion beam deposit in SXe excited at different wavelengths; (b) excitation spectrum for  candidate Ba$^+ $ lines.  This deposit was made and observed at 10~K.}
	\label{fig:Ba+_spectra}
\end{figure}

Emission spectra from a Ba$^+ $ ion deposit in SXe obtained with a tunable blue dye laser are shown in Fig.~\ref{fig:Ba+_spectra}(a).
Five new emission peaks at 532, 553, 568, 635 and 669~nm are observed.
Excitation spectra for these lines are shown in Fig.~\ref{fig:Ba+_spectra}(b).
The two strongest lines, 532~nm and 635~nm, have a similar excitation peak at 472~nm, indicating a common excited state and matrix site.
Two smaller peaks, 553~nm and 568~nm have a rise at the lowest wavelengths, suggesting an excitation peak at \textless 461~nm and perhaps also a common origin.
The other two peaks, 592~nm and 669~nm, exhibit increased excitation toward the shortest wavelengths probed.
Although the 592~nm peak is similar in wavelength to a strong emission peak of Ba in SXe, the very different excitation spectrum from that of Ba in Fig.~\ref{fig:blue_lines}(b) suggests a separate assignment.
The five lines that are not seen in the getter spectra, Fig.~\ref{fig:blue_lines} (a), and the 592~nm line are tentatively associated with Ba$^+ $.

\subsection{Bleaching}

\begin{figure}[h!tb]
	\includegraphics[width=0.45\textwidth]{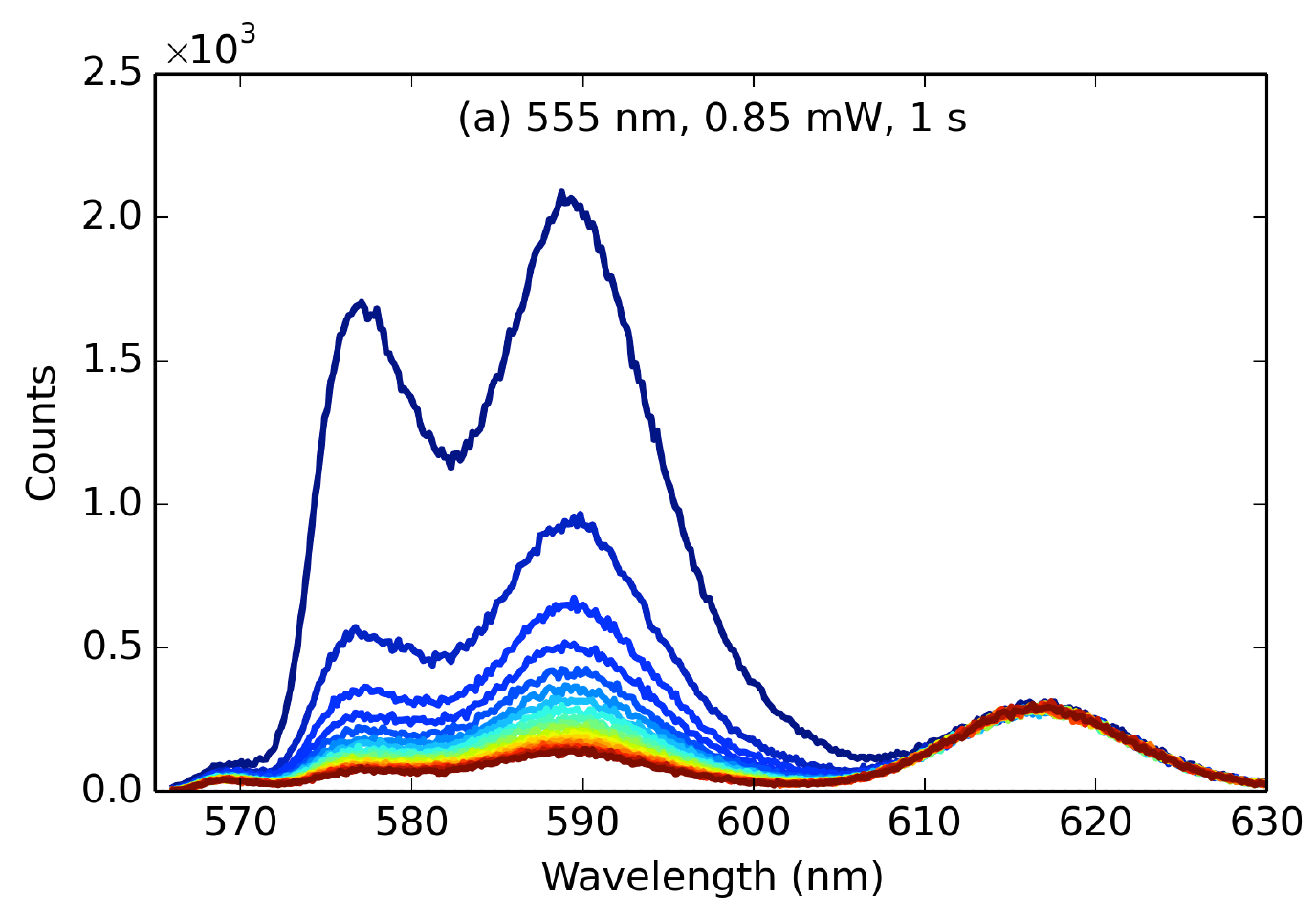}
	\includegraphics[width=0.45\textwidth]{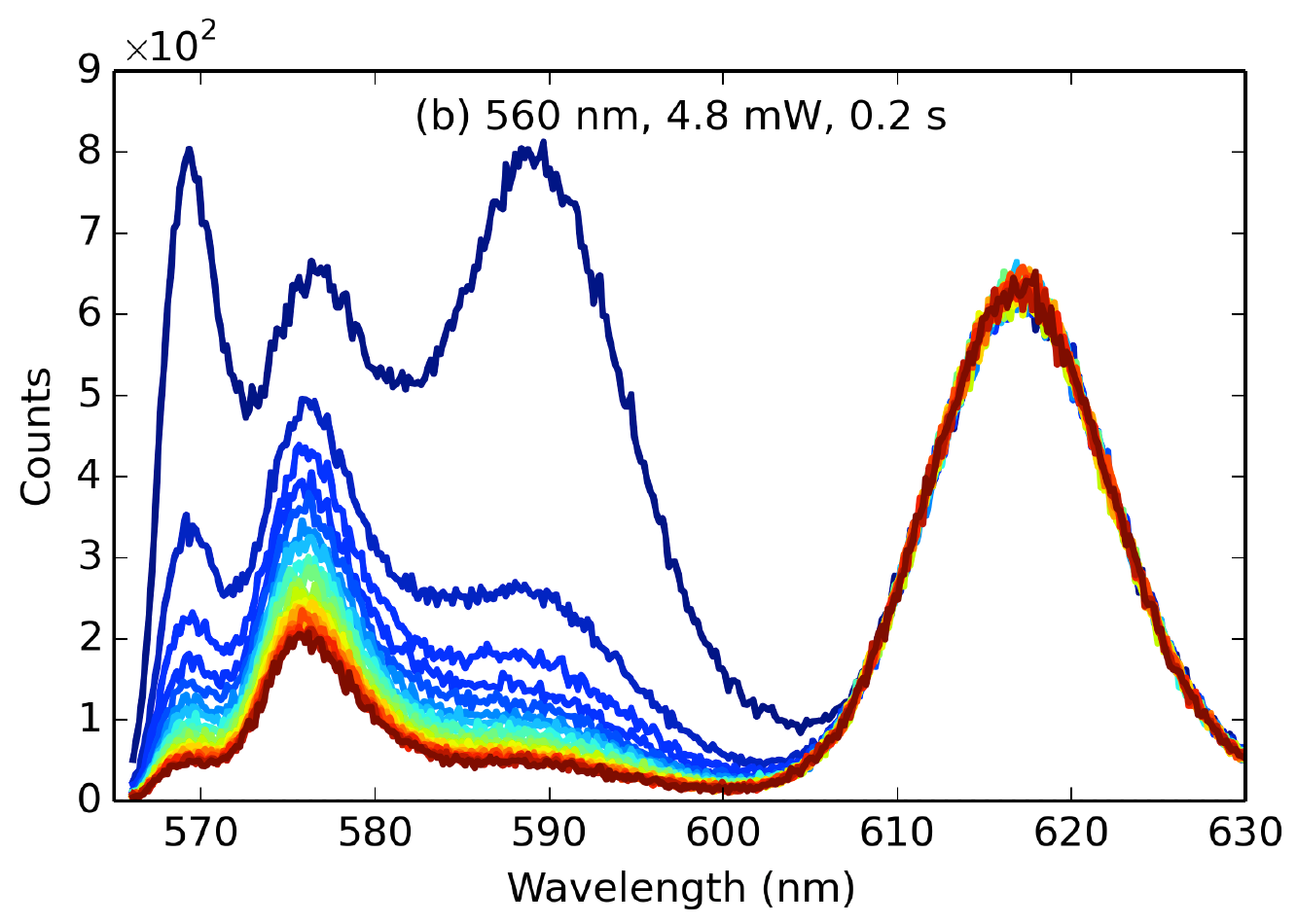}
	\includegraphics[width=0.45\textwidth]{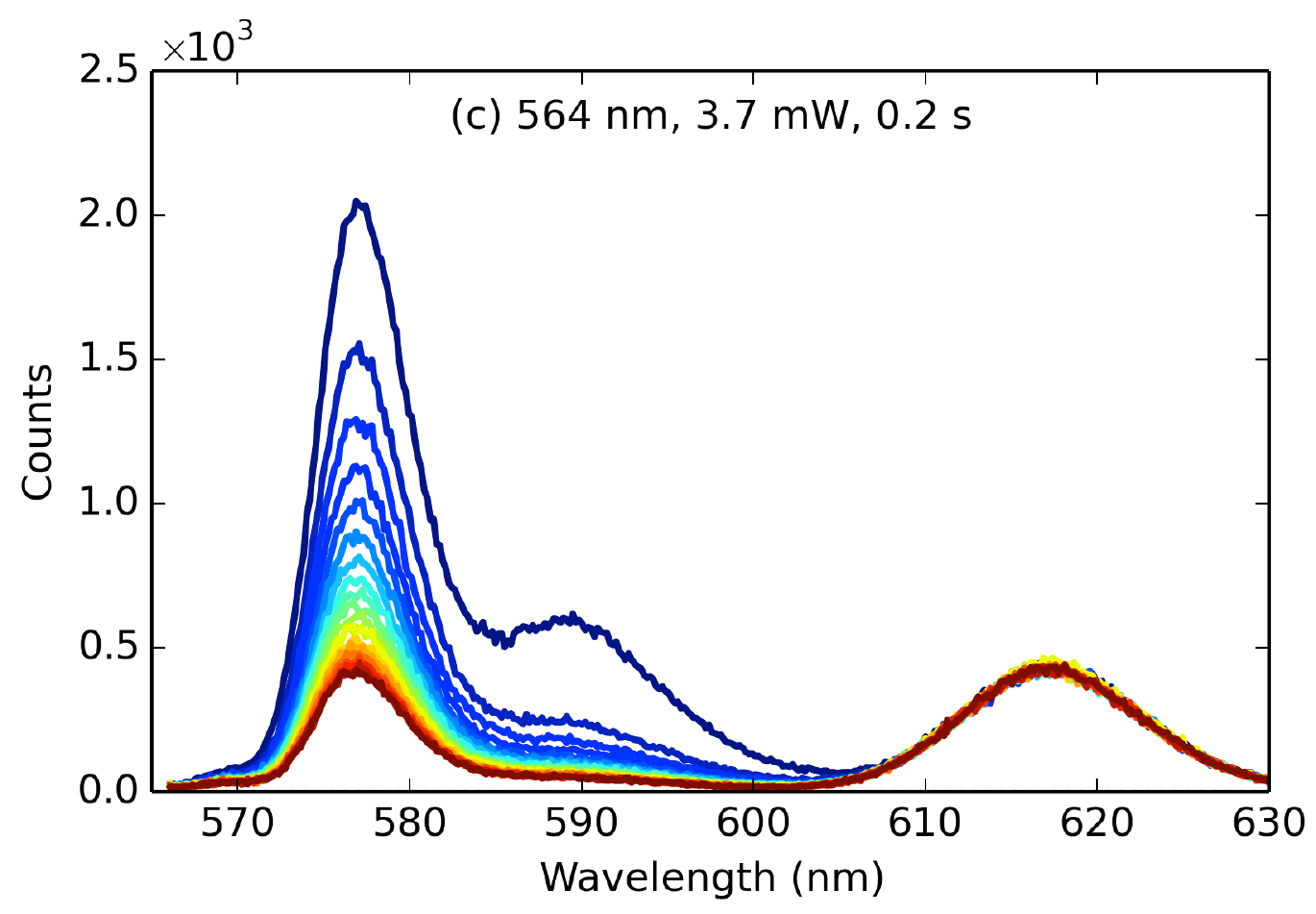}
	\caption{(color online) Bleaching of 568~nm, 575~nm and 590~nm fluorescence lines of Ba in SXe with (a) 555~nm, (b) 560~nm and (c) 564~nm excitation.  Every tenth exposure is shown starting with the second.  The colors range from blue (first) to red (last). The laser power and exposure time for each excitation wavelength are indicated. These are a Ba$^+$ deposits in SXe made at 45K and observed at 10~K.}
	\label{fig:bleaching_555nm}
\end{figure}

For many of the fluorescence peaks reported above, extended exposure of the matrix to intense laser light causes the fluorescence signal to decrease, or bleach, with time.
For example, successive spectra taken with a mildly focused dye laser beam at 555~nm, 560~nm and 564~nm with beam radius w=1000~$\mu$m and similar laser power-exposure time product are shown in Fig.~\ref{fig:bleaching_555nm}.
The peaks bleach at different rates depending on the excitation wavelength.
For example, with 555~nm excitation, the 575~nm emission peak bleaches more rapidly than the 590~nm peak.
On the other hand, with 560~nm excitation the 568~nm and 590~nm emission peaks bleach more rapidly than the 575~nm peak.
There are two additional peaks at 617~nm and 665~nm, that are initially weaker, but do not bleach.
The 665~nm peak is not shown because it was cut off by a filter in this data set.

In all the graphs in the previous sections, effort has been made to present spectra in which there had been minimal bleaching due to low laser power and an unfocused laser beam.

\subsection{Ba Molecular lines}

\begin{figure}[h!tb]
	\includegraphics[width=0.45\textwidth]{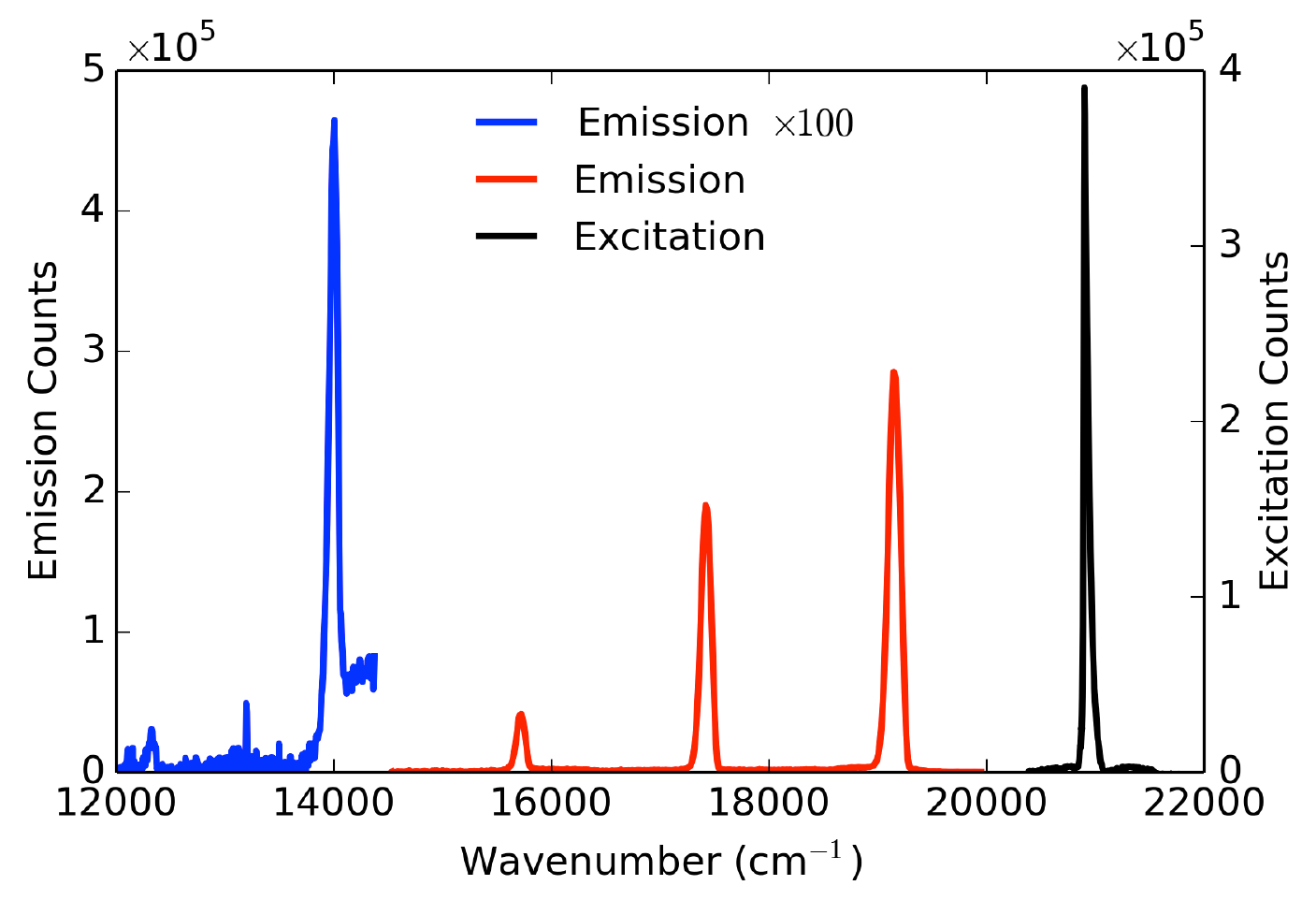}
	\caption{(color online) Excitation (black) and emission (red, blue) spectra with a set of sharp regularly-spaced lines observed in a Ba$^+$ deposit.}
	\label{fig:BaHx}
\end{figure}

Five additional emission lines observed only in Ba$^+ $ ion deposits were unusually narrow and had a common excitation spectrum with a sharp peak at 478.5~nm.  
The emission spectrum and the excitation spectrum for the 522~nm emission line are shown in Fig.~\ref{fig:BaHx}.  
The large and approximately constant spacing of 1723$\pm $48~cm$^{-1}$ suggests a vibrational sequence of a molecular electronic transition in a species with a small reduced mass, such as a hydrogen-containing molecule.  
The 478.5~nm excitation line is likely the electronic transition between v=0 vibrational states, whereas the five red-shifted fluorescence lines can be interpreted as transitions between the v$^\prime$=0 level of the excited state to the v$^{\prime\prime}$=1-5 levels of the ground state.  
The molecular vibrational constants in vacuum for several barium hydride molecules with electronic transitions in the blue region of the spectrum are given in Table~\ref{tab:BaH_table}.
All of these candidate molecules have ground state vibrational frequencies significantly smaller than the observed value.
The small matrix shift of the vibrational constant of the neutral species, BaH and BaH$_2$, in SAr argues against attributing the discrepancy to a large matrix shift of these species in SXe.
Nevertheless, the situation could be different for an ion.
For comparison, the vibrational constants of  H$_2$ and H$_2^+$ are also given.
These are much greater than the observed value.  
If there is H$_2$ impurity content in the matrix, an expected species is BaH$_2^+$ for Ba$^+ $ ion deposits.  
However, measured or calculated ground state vibrational constants for BaH$_2^+ $ were not found in the literature.

\begin{table}[h!bt]
 \begin{ruledtabular}

	\caption{Ground state vibrational constants $\omega_e$ in cm$^{-1}$ of simple barium and hydrogen molecules in vacuum and in solid argon matrix.}
	\label{tab:BaH_table}
	\begin{center}
	\begin{tabular}{c c c c}
	Species & Vacuum & SAr & ref\\
	\hline
	BaH$_2$ $\nu_3$ & 1102$^a$ & 1069 &\cite{Wang2004}\\
	BaH$_2$ $\nu_1$ & 1167$^a$ & 1129 &\cite{Wang2004}\\
	BaH &1168 & 1132 &\cite{Magg1988, *Wang2004}\\
	BaH$^{+}$ & 1370$^a$ & & \cite{Allouche1993}\\
	H$_2^{+}$ & 2324$^a$ & &\cite{Beckel1970} \\
	H$_2$ & 4400 & &\cite{Herzberg1959}\\
	H$_2$Xe & 4398 & &\cite{McKellar1971}\\
	BaH$_2^{+}$ & & &\\
	\hline
	Observed SXe &  1723(48) & &\\
	\hline
	$^a$theoretical
	\end{tabular}
	\end{center}
\end{ruledtabular}
\end{table}

\begin{figure}[h!tb]
	\includegraphics[width=0.45\textwidth]{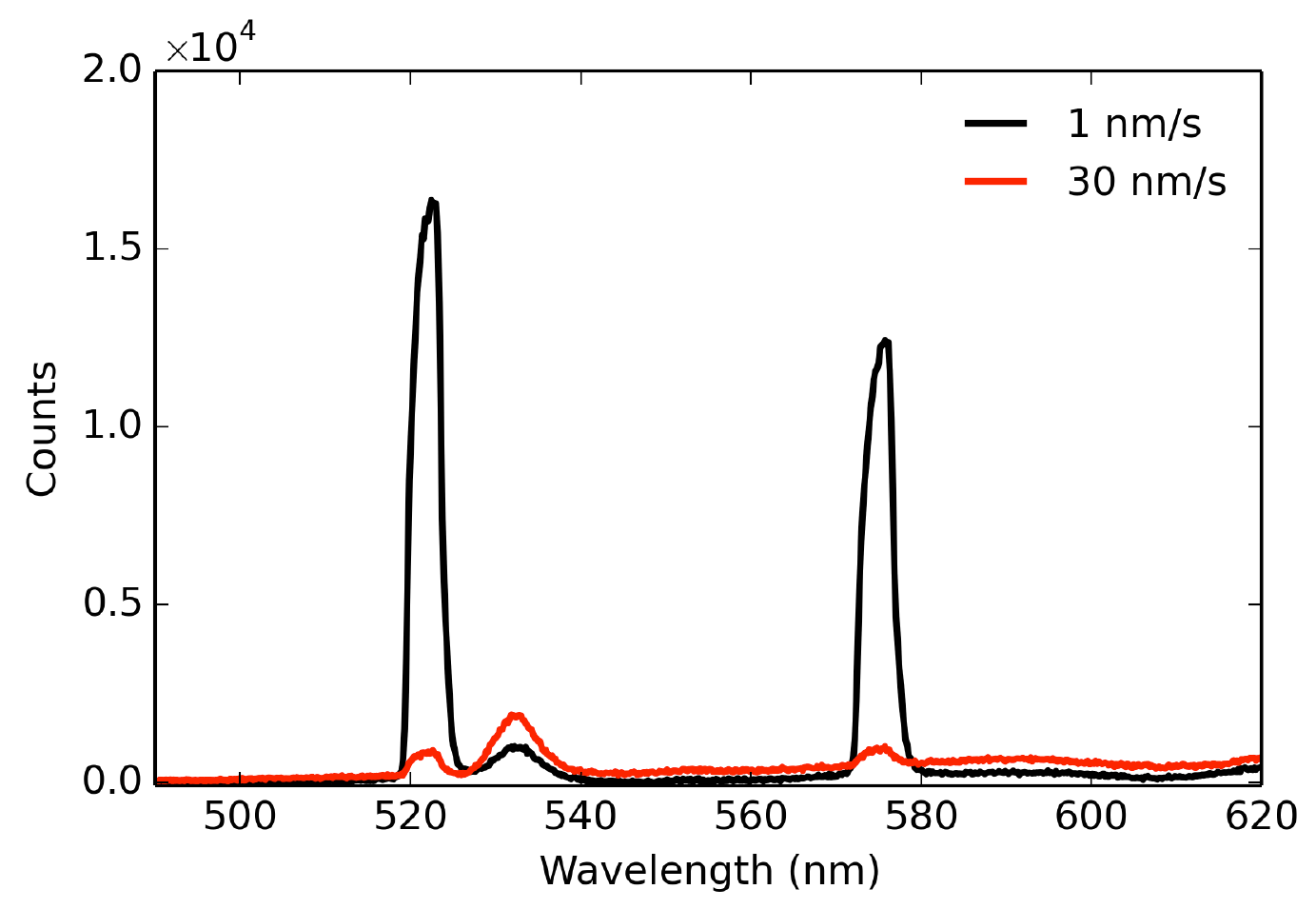}
	\caption{(color online) Emission spectra of a Ba$^+$ deposit in SXe made and observed at 10~K with 478.0~nm dye laser excitation at low (black) and high (red) xenon gas leak rate.  The matrix growth rate was 1 and 30~nm/s, respectively.}
	\label{fig:BaHx_leak_rate}
\end{figure}

Two experiments were performed to test the influence of the hydrogen impurity content in the matrix on the strength of these lines.
One way to vary the relative concentration of residual gas molecules in the matrix is to vary the Xe leak rate, as discussed above.
Emission spectra with 478.0~nm excitation at low leak rate (higher impurity concentration) and higher leak rate (lower impurity concentration) are shown in Fig.~\ref{fig:BaHx_leak_rate}.  
It is observed that the sharp emission peaks at 522~nm and 575~nm are much reduced in the purer matrix.  

\begin{figure}[h!tb]
	\includegraphics[width=0.45\textwidth]{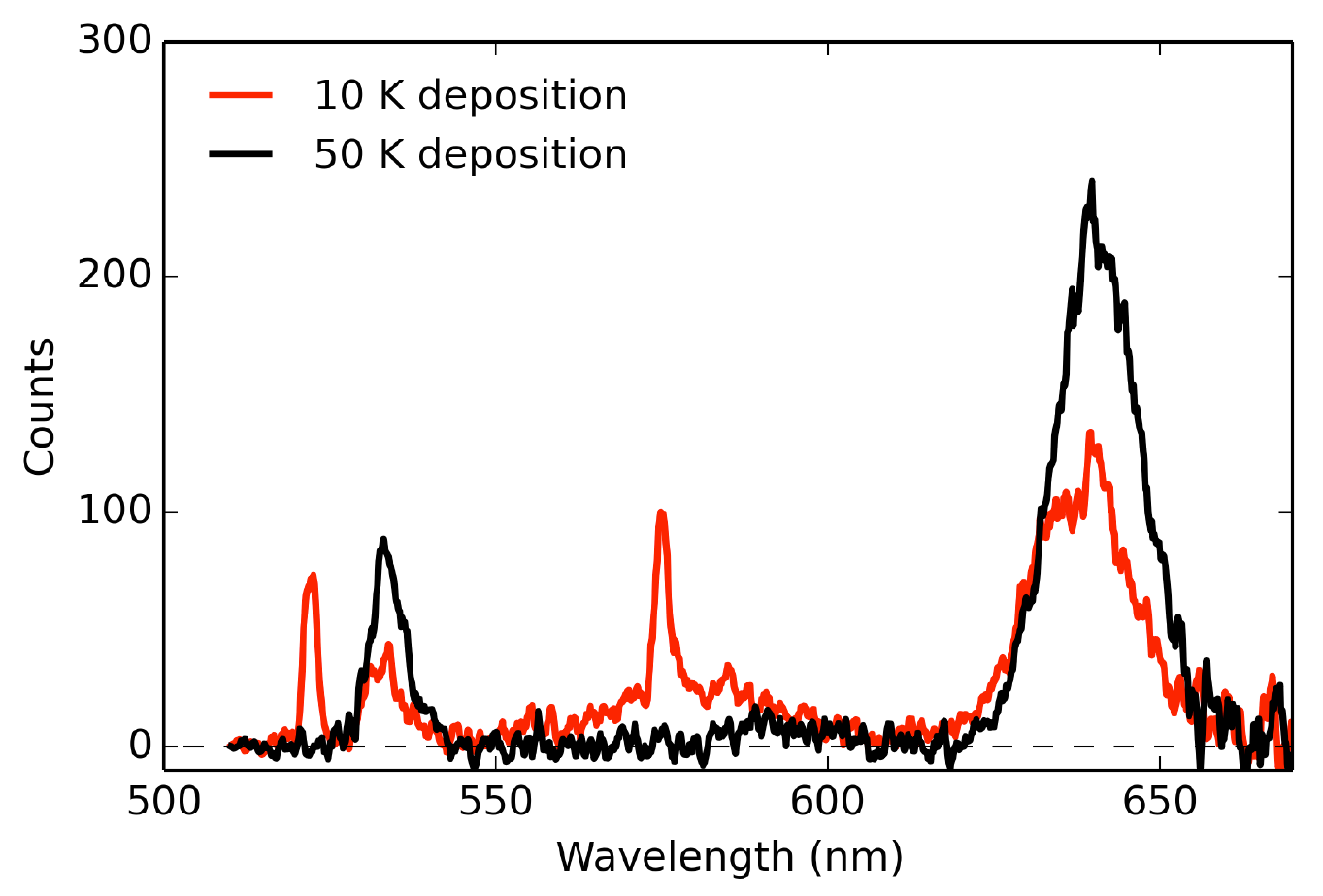}
	\caption{(color online) Fluorescence spectra of Ba$^+$ ion deposits in SXe made at two different temperatures and observed at 10~K.  The excitation wavelength was 478.1~nm.}
	\label{fig:BaHx_vs_T}
\end{figure}

In addition, one can independently control just the H$_2$ impurity content in the matrix through the temperature at which the matrix is deposited.  
When a pure residual gas matrix (minimal Xe) is heated, residual gas pressure measurements vs. window temperature reveal that H$_2$ evaporates at 11-20~K, whereas N$_2$, Ar and O$_2$ start evaporating at around 30-32~K. 
Thus a SXe matrix deposited at $\ge$ 25~K should contain much less H$_2$ than a matrix deposited at the usual 10~K.  
Spectra for SXe matrices deposited at 10~K and 50~K are shown in Fig.~\ref{fig:BaHx_vs_T}.  
The absence of the 522~nm and 575~nm lines in the higher temperature deposit further confirms the identification of these molecular lines with a barium hydride molecule of some sort, denoted below as BaH$_x^+$.

\section{Discussion}

\begin{figure}[h!tb]
	\begin{center}
	\includegraphics[width=0.45\textwidth]{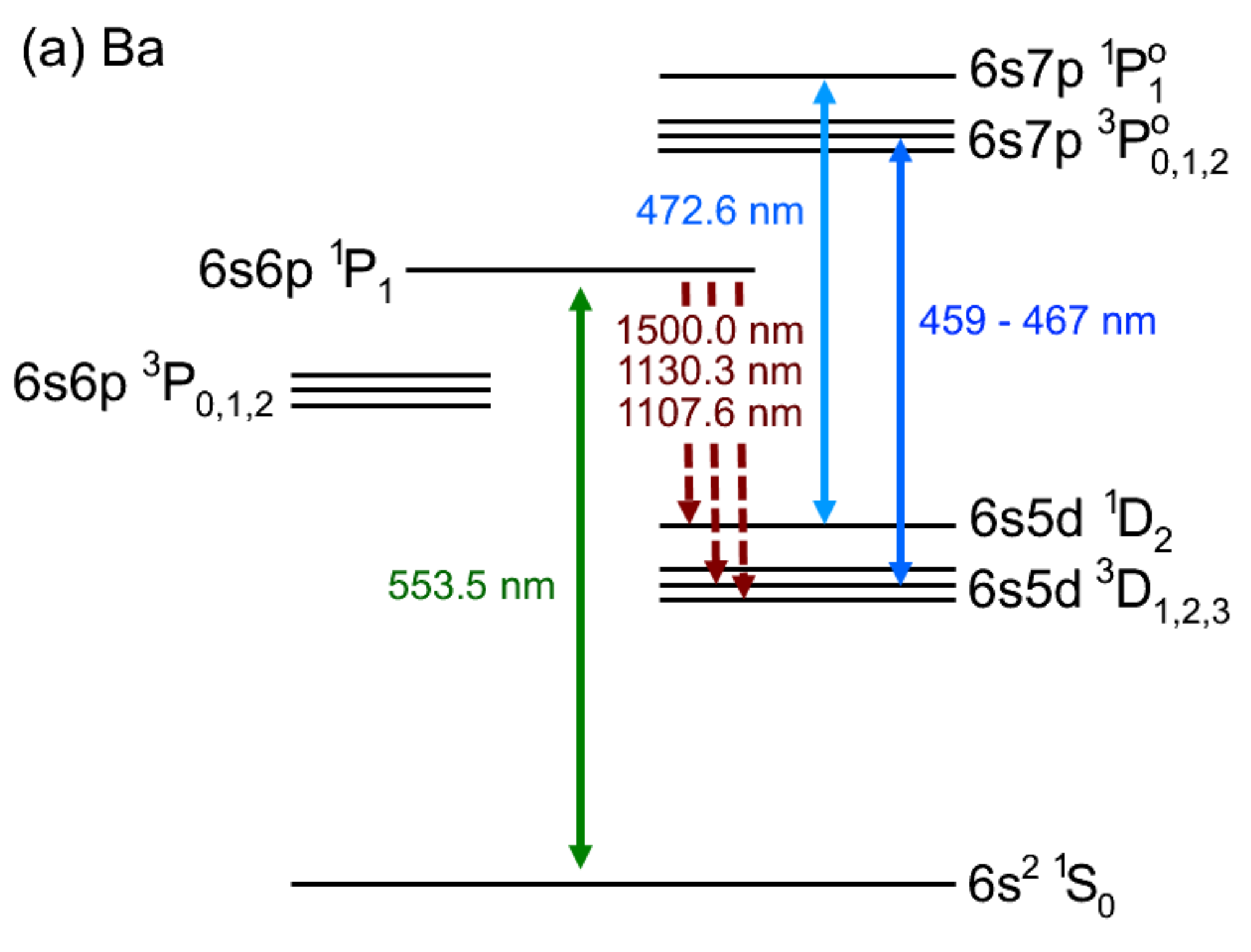}
	\includegraphics[width=0.45\textwidth]{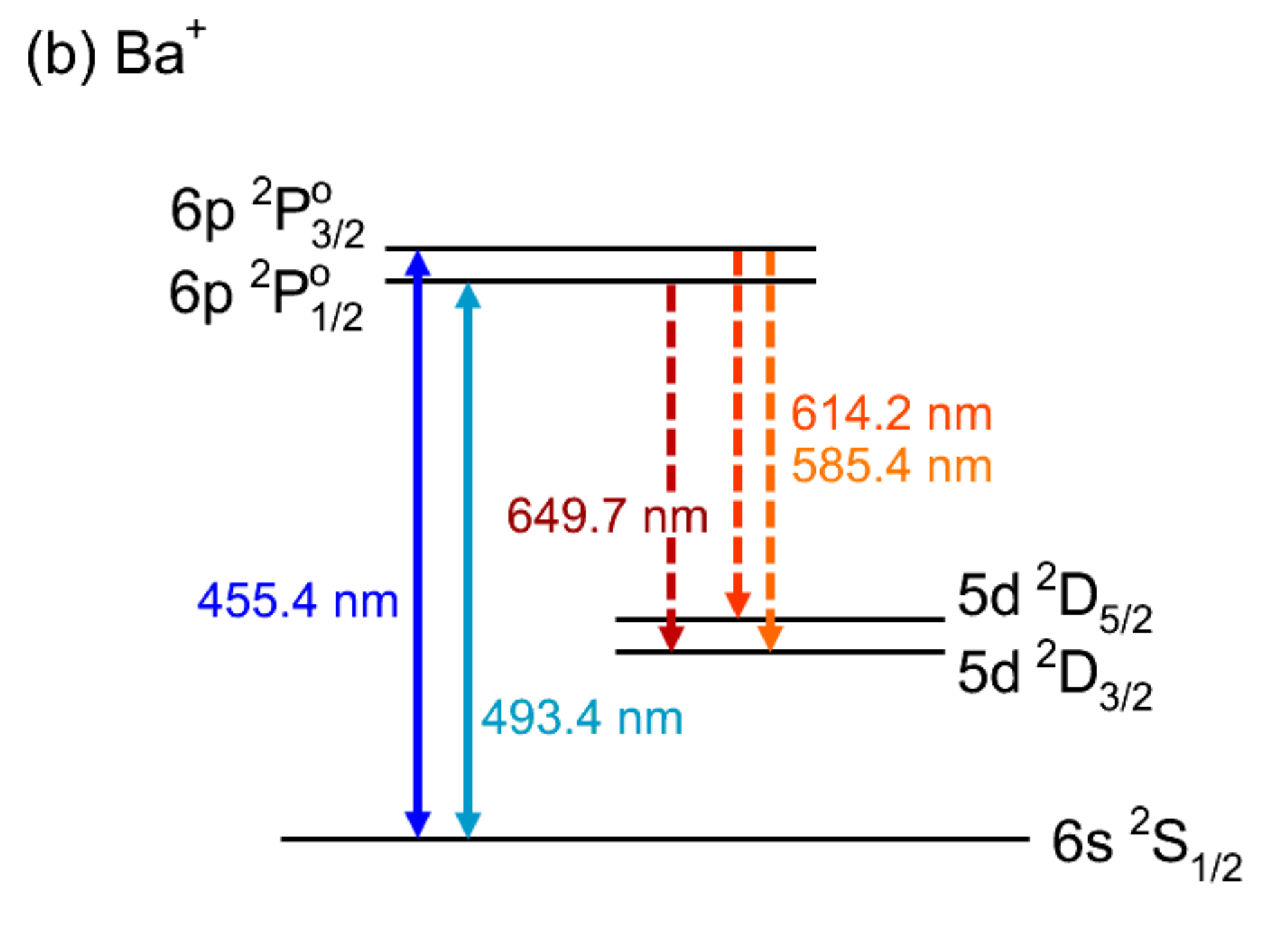}
	\caption{(color online) Energy levels of (a) Ba and (b) Ba$^+$ in vacum.}
	\label{fig:Ba_energy_levels}
	\end{center}
\end{figure}

The energy level diagram for the lowest lying levels of Ba in vacuum is shown in Fig.~\ref{fig:Ba_energy_levels}(a).  
From the ground $6s^2 ~^1S_0$ state, the strongest transition in vacuum is at 553.5~nm to the $6s6p ~^1P_1$ state.  
This state decays almost exclusively back to the ground state, but has weaker decays at around 1.1~$\mu$m and 1.5~$\mu$m to three $6s5d$ states with a branching ratio of 1/350.  
The strong absorption band observed around 558~nm and 530~nm for Ba atomic deposits in SXe and SAr, respectively, has been assigned to this transition.  
The red-shifted emission peaks at 565-590~nm and 530-540~nm, respectively, similarly are assigned to the return transition to the ground state in these matrices.

In vacuum, after about 350 cycles of this transition, a high fraction of the population accumulates in the $6s5d$ states, which have long radiative lifetimes, from 0.2~s for $^1D_2$ to 69~s for $^3D_2$.
To achieve a magneto-optical trap (MOT) in vacuum, three infrared repumping lasers were applied to overcome optical pumping and depopulate these long-lived metastable states \cite{De2009}.

In SXe, the lifetimes of these metastable states may not be as long.
There may exist a small odd parity component of the electric field experienced by the Ba atoms in various matrix sites due to the electron cloud of neighboring Xe atoms.
This could make the parity-forbidden decays of the 6s5d states to the $6s^2$ ground state weakly allowed by an admixture of wavefunctions of odd-parity states, e.g., $6s6p$ states.
It is not known how significant this effect will be.

In this model, if the $6s^2 \rightarrow 6s6p$ excitation rate is much greater than 10$^3$ times the $6s5d$ state decay rates, the populations of the 6s5d states can become comparable to that of the ground state.
This could be a rapid and dominant effect with focused resonant laser excitation.
Even with white light excitation, during lengthy absorption measurements, some optical pumping to metastable 6s5d state may occur.  
Thus one possible assignment of the weak absorption triplets observed in the blue for Ba in SXe and SAr could be absorptions out of these metastable states.
Indeed, transitions from these levels to the $6s7p$ states exist in vacuum in this wavelength region.
The 483~nm and 492~nm emission lines of Ba in SXe could be $6s7p \rightarrow 6s5d$ transitions.

An alternative explanation is matrix sites with large blue shifts on the $6s^2 \rightarrow 6s6p$ transition.
Such large shifts do occur for alkali atoms in noble gas matrices. 
For example, calculations have shown that the ``blue'' and ``violet'' triplets of Li, Na and K atoms in SAr matrices are due to atoms in sites with one and four neighboring vacancies, in which neighbor interactions with the alkali atom are relatively strong \cite{Jacquet2011}.
Further experimental and theoretical work is needed to assign these new blue transitions. 

The energy levels of Ba$^+$ in vacuum, shown in Fig.~\ref{fig:Ba_energy_levels}(b), are simpler than those of Ba.
The excitation peaks at 472~nm and  \textless 461~nm in Fig.~\ref{fig:Ba+_spectra}(b) could be the 6s-6p transitions.
The various emission peaks in Fig.~\ref{fig:Ba+_spectra}(a) could be associated with the 6p-5d transitions or red-shifted 6p-6s transitions.
Comparison spectra from other atomic ions in noble gas matrices are few, so it is not possible to make assignments based on general trends.
An extended excitation range with the blue dye laser, careful bleaching studies, pulsed laser excitation and theoretical modeling would help.

All observed lines and best assignments are summarized in Table~\ref{tab:fluorescence_table}.
Similar emission peaks obtained in SAr are also indicated.
For the spectra in SAr matrix, the dye laser was not used.
Thus the argon laser line with the strongest excitation is listed in parentheses in the excitation column.
The modest blue shift of each line in SAr relative to those in SXe is consistent with expectations from previous matrix isolation spectroscopy results \cite{Crepin1999}.

\begin{table*}[t]
 \begin{ruledtabular}

	\caption{Summary of peak wavelengths of observed absorption and emission lines in SXe and SAr, species and transition assigned and vacuum wavelengths for comparison.  Excitation wavelengths in parentheses are the argon ion wavelengths that give maximum emission.}
	\label{tab:fluorescence_table}
	\begin{center}
	\begin{tabular}{c c c c c c c}
	Species & Transition & Vacuum (nm) & Excitation (nm) & Emission (nm) & Excitation (nm) & Emission (nm)\\
	\hline
	\multicolumn{3}{c}{} & \multicolumn{2}{c}{SXe} & \multicolumn{2}{c}{SAr}\\
	\hline
	Ba & & & 460 & 484 & &\\
	Ba & & & 467 & 492 & (454-473) & 486\\
	Ba & $6s^2 ~{^1}S_0\leftrightarrow 6s6p ~{^1}P_1$ & 553.5 & 559 & 568& & \\
	Ba & $6s^2 ~{^1}S_0\leftrightarrow 6s6p ~{^1}P_1$ & 553.5 & 564 & 575&  & \\
	Ba & $6s^2 ~{^1}S_0\leftrightarrow 6s6p ~{^1}P_1$ & 553.5 & 560 & 590 & 532 & 550\\
	Ba & & & \textgreater 564 & 617 & &\\
	Ba$^{+}$ & & & 472 & 532 & (454-473) & 500\\
	Ba$^{+}$ & & &  \textless 461 & 553 & (473-496) & 540\\
	Ba$^{+}$ & & &  \textless 461 & 568\\
	Ba$^{+}$ & & &  \textless 461 & 592\\
	Ba$^{+}$ & & & 472 & 635 & (488) & 592\\
	Ba$^{+}$ & & &  \textless 461 & 669\\
	BaH$_x^{+}$ & v'=0$ \rightarrow$ v"=1 & & 478 & 522 & (473) & 516\\
	BaH$_x^{+}$ & v'=0 $\rightarrow$ v"=2 & & 478 & 574 & (473) & 566\\
	BaH$_x^{+}$ & v'=0 $\rightarrow$ v"=3 & & 478 & 637\\
	BaH$_x^{+}$ & v'=0 $\rightarrow$ v"=4 & & 478 & 712\\
	BaH$_x^{+}$ & v'=0 $\rightarrow$ v"=5 & & 478 & 814\\
	\end{tabular}
	\end{center}

\end{ruledtabular}
\end{table*}

To progress towards the goal of single atom or ion detection, it is important to determine the fluorescence rate per atom from the measured fluorescence count rate.
This requires a knowledge of the number of Ba atoms per area in the matrix deposit.
For getter deposits, the Ba flux is not known, but the Ba atom density per area in the matrix may be determined from absorbance measurements.
The absorbance A$_\lambda$ is related to the Ba density per volume $N$ and absorption cross-section $\sigma (\lambda)$ by 
\begin{equation}
	A_\lambda=\sigma(\lambda) N \ell
\end{equation}
where $l$ is the path length through the Ba layer in the matrix.  
The quantity $N\ell$ is the Ba density per area.  
The number of Ba atoms in a laser beam of area $a$, $N\ell a$, can be determined for a given deposit if the absorbance is measured and the absorption cross section is known.

A reasonable expectation is that the integrated cross section for a strong transition of a Ba atom in vacuum is conserved in the matrix.
This is an assumption of conservation of oscillator strength.
In this approximation, the cross section $ \sigma(\lambda) $ for a Ba atom in the SXe matrix (averaging over different matrix sites) may be estimated using the shape of the absorption curve in Fig.~\ref{Fig:BaAbsFlur}(a).
\begin{equation}
	\sigma(\lambda)=A_\lambda\frac{\int\sigma(\lambda)d\lambda}{\int A_\lambda d\lambda}
\end{equation}
The peak absorption cross section obtained is  $\sigma$($\lambda_0$)=1.2$\times$10$^{-15}$~cm$^2$ at $\lambda_0$=558~nm.  
Using equation (1) the area density of Ba atoms for the data set in Fig.~\ref{Fig:BaAbsFlur} can be estimated as 1$\times$10$^{13}$~atoms/cm$^2$ 
for this particular deposit.

For large Ba deposits, where the Ba area density is known from absorption measurements, the expected fluorescence rate of transitions to the ground state, $F$, in photons/s/atom can be predicted.
In the absence of non-radiative transitions from the excited state, it is given by
\begin{equation}
F = W_{12} \epsilon _b
\end{equation}
where $W_{12}$ is the excitation rate and $\epsilon_b$ is the branching ratio to the ground state.
In terms of the laser intensity $I$ and photon energy $h\nu$, $W_{12}$ is given by
\begin{equation}
W_{12}= \sigma(\lambda) I/h\nu
\end{equation}
The observed fluorescence rate, $F_{obs}$, can be calculated from the measured fluorescence count rate and the estimated photon detection efficiency, $\epsilon_d$, in counts per photon emitted.
The ratio of observed to predicted fluorescence rate is the fluorescence quantum efficiency of the transition,
\begin{equation}
\epsilon_{QE}=F_{obs}/F
\end{equation}
In vacuum, $\epsilon_{QE}$ is 1, but in the matrix it could be less than 1 due to non-radiative transitions, for example.
Typical numbers determined from large Ba deposits in SXe gave $\epsilon_{QE}$ on the order of 0.1-1~\%.  These results should be taken as lower limit estimates, as bleaching effects had not been studied carefully at the time of these early experiments.

For ion deposits, the area density of Ba$^+$ ions co-deposited with the matrix is known from the ion current and the duration of the deposit.  
However, the fraction that neutralizes to Ba has not yet been determined.  
Thus the quantum efficiency for Ba atom fluorescence determined from ion experiments, typically 0.01-1 \%, is also only a lower limit.

For single Ba atom detection with barium tagging applications in mind, it is important to understand this loss of efficiency in fluorescence detected.
As a comparative benchmark, single fluorescent molecules, that can be imaged routinely in a variety of environments, and even have been imaged in SXe in one case (\cite{Sepiol1999}), typically have fluorescence quantum efficiencies in the 10's of percent range.

Initial studies of the bleaching of the primary Ba atom fluorescence at 568~nm, 575~nm and 590~nm  with attenuated laser beams and defocussing to various laser beam diameters have been performed.
To assure uniform intensity in the region of interest, for comparison to a simple rate equations model, the Ba fluorescence was collected only from a central area of the laser beam where the intensity is within 90\% of the maximum.

Normalized experimental bleaching curves have been compared to a rate equations model with one excitation rate W$_{12}$ and 7 spontaneous emission rates A$_{ij}$ between the five states of interest shown in Fig.~\ref{fig:Ba_energy_levels}(a).  Agreement is relatively good using values for these parameters close to the W$_{12}$ value calculated as discussed above and the free atom A$_{ij}$ rates.  In some cases the shape of the decay curve is matched better by adding an additional loss rate from the $6s6p~^1P_1$ state, up to 10$^5$~s$^{-1}$, that may, for example, account for a change in matrix site as a result of an excitation cycle.

The addition to the model of rapid non-radiative decays out of the $6s6p ^1P_1$ state at rates of $10^{10}-10^{12}$~s$^{-1}$ could account for low $\epsilon_{QE}$.
The bleaching curves of the model would be unaffected if these rapid non-radiative decays occurred only to the ground $6s^2$ state.
However, this is a physically improbable scenario.
The most likely final state for such non-radiative decays is the nearby $6s6p ~^3P_2$ state, as has been found with Ba in low pressure Xe gas and with Hg in SAr matrix \cite{De2009,*Crepin1994}.
Since this state undergoes strong radiative decay to the metastable $6s5d ~^3D$ states, this leads to very rapid optical pumping in the model and complete disagreement with observed bleaching curves.

Recent theoretical work opens the possibility of a deeper understanding of the transitions associated with the various observed peaks, as well as the observed bleaching rates.
Alkali atoms Li, Na and K in solid Ar, Kr and Xe have been modeled using a method based on alkali-noble gas pair potentials \cite{Ryan2009}.
Matrix sites of alkali atoms upon deposition were found by molecular dynamics simulations.
Absorption triplets could be assigned to specific matrix sites, but emission spectra had poor agreement with experiment.
Using a model based on core polarization pseudopotentials, a much better match to experimental emission spectra was obtained for Na in SAr, and various peaks in observed spectra could be assigned to four distinct matrix sites \cite{Jacquet2011}.
Recently, the molecular pair potentials of various states of BaXe and BaXe$^+$ have been calculated \cite{Abdessalem2013}.
Thus detailed simulation of the spectra of Ba and Ba$^+$ in SXe matrix sites could be done.
This opens the possibility of making assignment of the unidentified lines in Table~\ref{tab:fluorescence_table} and understanding the bleaching and quantum efficiency of various transitions through theoretical comparisons.

For example, the recent calculations indicate that the 2$^1\pi$ molecular potential curve of BaXe, corresponding to the $6s6p~^1P_1$ atomic state of Ba, is crossed near its minimum by the potential curve of the repulsive 2$^3\Sigma^+$ state arising from the 6s6p${^3}$P atomic states of Ba.
Population transfer at this level crossing is spin-forbidden, but might occur weakly.
This could explain the non-radiative transition, $6s6p~^1P_1\rightarrow 6s6p~^3P_2$, observed in gas phase collisions.
Level crossing may also contribute to the loss of quantum efficiency of Ba in SXe found in this work, although additional non-radiative pathways would be required have consistency with the bleaching data.

In BaXe$^+$, the equilibrium radii for the ground X$^2\Sigma^+$ state and the 2$^2\pi$ state are calculated to be nearly identical, as both bound states experience the same dominant charge-induced dipole potential term.
Thus, in this molecule, the low temperature absorption and emission spectra corresponding to the primary transitions of Ba$^+$ in vacuum at 455.4~nm and 493.4~nm are expected to be quite sharp, with a Stokes shift of only a few nm.
If similar considerations apply for Ba$^+$ in SXe, the emission on these transitions could have been missed because it was cut off by the Raman filters used in this work.
The BaXe$^+$ transitions to the states arising from the 5d atomic configuration, 1$^2\Delta$, 1$^2\pi$ and 1$^2\Sigma^+$, would be in the yellow-green region and sharp for the first two bound states and in the red region and broader for the third weakly bound state with a larger repulsive term.
These are candidate assignments for the observed emission lines in SXe that have been tentatively listed as Ba$^+$.

\section{Imaging Ba atoms}

\begin{figure*}[h!tb]
	\begin{center}
	\includegraphics[width=1\textwidth]{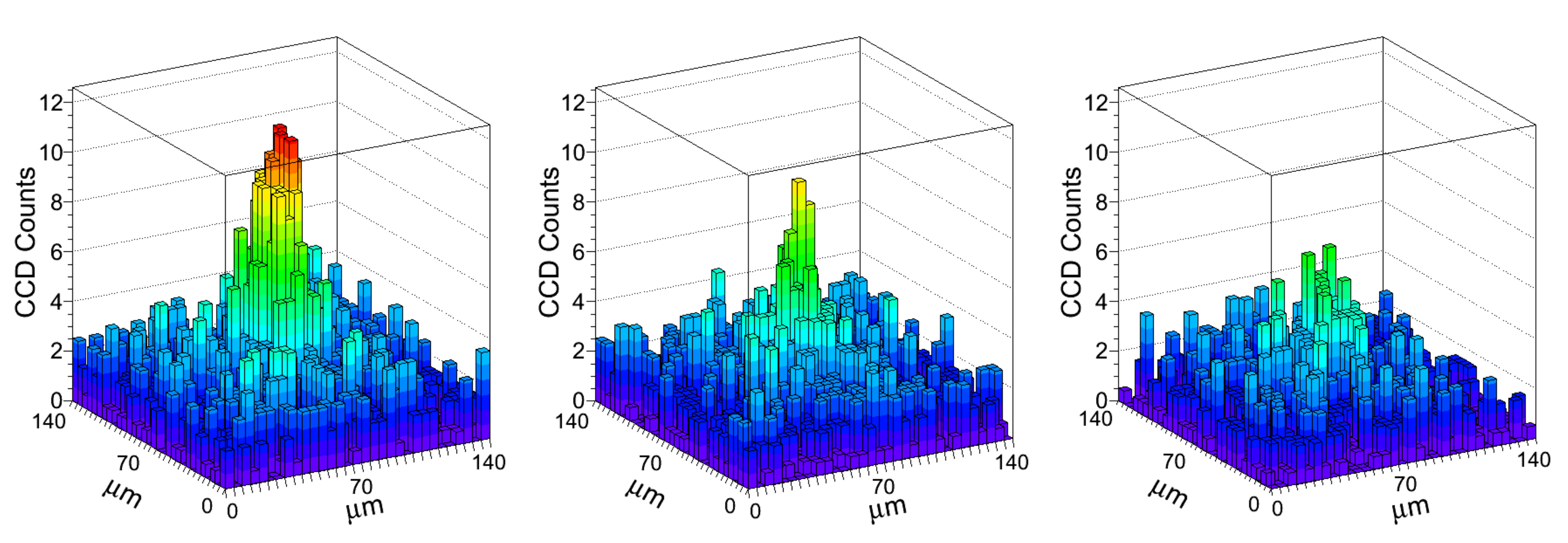}
	\caption{(color online) Successive 1~s images of a small number of Ba atoms from about $10^4$ Ba$^+$ ions deposited within the laser beam area.  The excitation laser was at 564~nm.}
	\label{fig:Ba_image}
	\end{center}
\end{figure*}

A series of three consecutive images of a small number of Ba atoms in a focused dye laser beam is shown in Fig.~\ref{fig:Ba_image}.
They were taken with a modified optical system in which a f/0.9 aspherical collection lens was used for higher light collection efficiency, and the zoom lens projected the image directly on the CCD chip.
This avoided angular and aberration limitations, as well as losses, of the spectrometer in zero order imaging mode.
A optical bandpass filter with sharp cutoffs passed 578-603~nm emission.
In this case about $10^4$ Ba$^+$ ions were deposited within the 1/e radius of the laser beam (1/e$^2$ radius $w$=2.3~$\mu$m), and an unknown percentage of them neutralized to Ba atoms.
In the first 1~s frame, a sharp peak containing about 500 counts is observed.  
Due to bleaching, the fluorescence peak is reduced in the second frame, and is nearly gone by the third frame.
The laser intensity was attenuated by 10$^3$ in this experiment, to the $\mu$W level, in order to observe the decay due to bleaching with 1~s frames of the CCD camera.

With an optical magnification of 4.3, the effective pixel size in this image is 4.7~$\mu$m.
The FWHM of the image, 30$\times$43~$\mu$m, is an order of magnitude larger than the dye laser FWHM.
This blurring of the image is mainly due a rough SXe surface in this experiment.

One of the remarkable observations in this work is that after a sample with a large Ba or Ba$^+$ deposit in SXe is evaporated by heating to 100~K, there is no evidence of any remaining Ba or Ba$^+$ fluorescence in the next deposit at the current level of sensitivity.
It is not known whether the  Ba or Ba$^+$ evaporated with the xenon or if it remained on the window as a non-fluorescing metallic or oxide coating, for example.
This lack of a ``history'' effect is potentially an important advantage for Ba tagging in a double beta decay experiment.

A near-term goal of this barium tagging research is to obtain images of single Ba atoms or Ba$^+$ ions in SXe on a cold sapphire window.
The main factors that limit single atom or ion imaging in the current setup are (1) limited fluorescence duration due to bleaching, (2) low quantum efficiency, (3) modest photon detection efficiency and (4) optical system aberrations.
Improvement in photon detection efficiency and perfecting optical system image quality to the diffraction limit are straightforward through the implementation of high collection efficiency microscope objectives commonly used in single molecule imaging.
Greater care is needed to ensure conditions of smooth matrix surfaces.
Work is underway to study the first two factors, which involve the fundamental atomic physics of barium atoms and ions in SXe matrices.
If bleaching can be overcome, higher intensities by 10$^3$ or more and longer exposures would yield single atom signals comparable to the image in Fig.~\ref{fig:Ba_image}.

\section{Conclusions}

Strong and stable fluorescence of Ba atoms in solid xenon matrices at low temperature has been demonstrated.
Absorption and fluorescence spectra associated with the $6s^2 \!\! \rightarrow \!\! 6s6p$ transition are reported.
Excitation spectra and temperature and bleaching dependences demonstrate the existence of different matrix sites for the Ba atoms in solid xenon matrix.
Corresponding spectral features are found for Ba in solid argon matrix, somewhat blue-shifted relative to SXe.
Additional absorption and fluorescence peaks of Ba in SXe may represent transitions out of the metastable $6s5d$ states of Ba or blue bands associated with the primary transition in different matrix sites.

The strong neutral Ba absorption and emission peaks are also found in Ba$^+$ deposits into SXe, indicating the occurrence of some neutralization.
In addition, sharp peaks associated with well-resolved vibrational lines of an electronic transition in a yet unidentified barium and hydrogen species are found.
Additional peaks not found in Ba deposits are tentatively assigned to Ba$^+$ transitions in SXe.  
Further experimental and theoretical studies are needed to fully understand all the observed spectral features.

In the current work, bleaching, fluorescence quantum efficiency and photon detection efficiency limit Ba atom imaging capability to $\le$10$^4$ atoms.
Optics improvements, the addition of repumping lasers to overcome bleaching and a better understanding could make single Ba atom imaging possible, leading to a method for barium tagging in liquid xenon for nEXO.

\section*{Acknowledgements}

This material is based upon work supported by the National Science Foundation under Grant Nunber PHY-1132428 and the U.S. Department of Energy, Office of Science, Office of High Energy Physics under Award Number DE-FG02-03ER41255.

\bibliography{references10}
\end{document}